\documentclass[twocolumn,groupedaddress,superscriptaddress,amsmath,amssymb,pra,aps]{revtex4-2}

\usepackage{amsmath}
\usepackage{tabularx}
\usepackage{color}
\usepackage{tikz}
\usetikzlibrary{quantikz2}
\usepackage{graphicx, import}
\usepackage{subcaption}
\usepackage{hyperref}
\usepackage{parskip} 
\usepackage{booktabs}
\usepackage{multirow}
\usepackage{soul}
\usepackage[normalem]{ulem}

\usepackage{algorithm}
\usepackage{algpseudocode}
\usepackage{placeins}
\usepackage{float}
\usepackage{microtype}

\usepackage{xcolor}

\begin{document}

\title{Quantum-enhanced Markov Chain Monte Carlo for Combinatorial Optimization}
\author{Kate V.~Marshall}
\email{kate.marshall@ibm.com}
\affiliation{IBM Quantum, IBM Research Europe -- Hursley, United Kingdom.}
\author{Daniel J.~Egger}
\affiliation{IBM Quantum, IBM Research Europe -- Zurich, Switzerland}
\author{Michael Garn}
\affiliation{The Hartree Centre, STFC, Sci-Tech Daresbury, United Kingdom}
\author{Francesca Schiavello}
\affiliation{The Hartree Centre, STFC, Sci-Tech Daresbury, United Kingdom}
\affiliation{The Open Quantum Institute, Geneva, Switzerland}
\author{Sebastian Brandhofer}
\affiliation{IBM Quantum, IBM Research Europe -- Böblingen, Germany} 
\author{Christa Zoufal}
\affiliation{IBM Quantum, IBM Research Europe -- Zurich, Switzerland}
\author{Stefan Woerner}
\affiliation{IBM Quantum, IBM Research Europe -- Zurich, Switzerland}

\date{\today}

\newcommand{\kate}[1]{\textbf{\color{magenta}[Kate: #1]}}
\newcommand{\daniel}[1]{\textbf{\color{orange}[Daniel: #1]}}
\newcommand{\michael}[1]{\textbf{\color{blue}[Michael: #1]}}
\newcommand{\sebastian}[1]{\textbf{\color{teal}[Sebastian: #1]}}
\newcommand{\christa}[1]{\textbf{\color{violet}[Christa: #1]}}
\newcommand{\stefan}[1]{\textbf{\color{red}[Stefan: #1]}}

\begin{abstract}

Quantum computing offers an alternative paradigm for addressing combinatorial optimization problems compared to classical methods. However, despite recent hardware improvements, the execution of empirical quantum optimization experiments at scales known to be hard for state-of-the-art classical solvers is not yet in reach. In this work, we offer a different way to approach combinatorial optimization with near-term quantum computing. Motivated by the promising results observed in using quantum-enhanced Markov chain Monte Carlo (QeMCMC) for approximating complicated probability distributions, we combine ideas of sampling from the device with QeMCMC together with warm-starting and parallel tempering, in the context of combinatorial optimization. We demonstrate empirically that our algorithm recovers the global optima for instances of the Maximum Independent Set problem (MIS) up to 117 decision variables using 117 qubits on IBM quantum hardware. We observe an indication of more favorable scaling relative to the tested classical MCMC methods on the chosen MIS instances. MIS is practically relevant across domains such as financial services and molecular biology. Although some MIS instances with a few hundred variables can be difficult for classical solvers, the instances studied here are solved efficiently by CPLEX. 

\end{abstract}

\maketitle


\section{Introduction} \label{sec:introduction}

Combinatorial optimization is an important application for quantum computing, inspired by the natural formulation of such problems to finding the ground state of Ising Hamiltonians \cite{Abbas_Challenges_Opportunities_Opt_2024}. This mapping has motivated the search for quantum optimization algorithms that could offer super-polynomial speedups over classical approaches \cite{Farhi_2001, Farhi_2015, Hastings_2018, Basso_2022, Dalzell_2023, Shaydulin_2024}. There exist various approaches to studying quantum optimization, ranging from provably exact to heuristic algorithms. 
Heuristic approaches usually lack provable convergence guarantees, runtime guarantees, or both \cite{Abbas_Challenges_Opportunities_Opt_2024}. Nevertheless, quantum algorithms could outperform classical solvers for select problem instances. In the field of classical optimization, most practically used algorithms are heuristic. One such quantum heuristic is the Quantum Approximate Optimization Algorithm (QAOA) \cite{Farhi_2014}, which benefits from bounded worst-case performance guarantees, e.g., for MAXCUT on 3-regular graphs \cite{Wurtz_2021}. 
However, the search for a quantum advantage has proved extremely challenging, largely due to noise that continues to limit the width and depth of circuits executable on available hardware. 

Whilst we navigate these challenges, heuristic quantum-classical frameworks offer a practically motivated direction for near-term quantum computing research \cite{Runtime_Adv_Chandarana_2025, Eisert_AdvReview_2025}. For quantum optimization, hybrid quantum-classical algorithms and techniques such as warm-starting leave much to be hopeful about, even before fault-tolerance \cite{ReduMIS_2025, Abbas_Challenges_Opportunities_Opt_2024, Eisert_AdvReview_2025}. Today’s quantum devices--despite inherent noise--have achieved a scale, speed, and fidelity that enable computations beyond the reach of exact classical simulation \cite{Kim_2023_Utility, QuantumAdvantage1, QuantumAdvantage2}. Hence, we can now benchmark the capabilities of quantum-enhanced solvers on non-trivial optimization problem instances, such as the ones provided in the Quantum Optimization Benchmarking Library (QOBLIB) \cite{QOBLIB_Koch_2025}. This resource provides combinatorial optimization problem instances, benchmarking metrics, baseline results to assist the community in exploring the path to quantum advantage in optimization, as well as a foundation to better understand the potential practical usefulness of quantum optimization algorithms in addressing industry-relevant applications.

As quantum hardware becomes increasingly capable, we can begin to investigate theoretical results with non-trivial problem instances. One set of interesting theoretical and early numerical results include those from the quantum-enhanced Markov chain Monte Carlo (QeMCMC) algorithm, originally developed for approximating complex probability distributions \cite{QeMCMC_Mazzola_2021, QeMCMC_Layden_2023}. The authors in Ref.~\cite{QeMCMC_Layden_2023} showed that with a quantum proposal distribution, QeMCMC achieved up to a quartic speedup in reaching the target distribution over comparable classical Markov chain Monte Carlo (MCMC) methods. Subsequent work has investigated the behavior, limitations and variants of QeMCMC \cite{orfi2024barriers, orfi2024bounding, ferguson2025quantum, christmann2025quantum, chang2025quantum, Counterdiabatic_Q_Sampling_2025}. 

While asymptotic speedups for quantum optimization algorithms are only expected with fault-tolerant quantum devices \cite{Future_Bravyi_2022}, heuristic quantum optimization algorithms may demonstrate promising results in the near-term for select problem instances. Our novel contribution is to combine independent ideas from QeMCMC, warm-starting quantum optimization, and parallel tempering, to solve combinatorial optimization problems. We show the first realization of a practical end-to-end QeMCMC workflow on system sizes of more than 100 qubits. We then approach and solve instances of Maximum Independent Sets (MIS) to optimality, for increasing instance sizes up to 117 decision variables using IBM Quantum superconducting quantum computers. Instances of this problem class are already challenging to solve with state-of-the-art classical solvers with only a few hundred variables, in some cases \cite{QOBLIB_Koch_2025}. We observe an indication of more favorable scaling of our method in terms of increasing problem size, compared to similar classical MCMC implementations for the five selected problem instances studied. Related approaches have previously explored iterative updates to quantum optimization procedures, including longitudinal bias fields in quantum annealing \cite{Grass2019} and adaptive bias fields in QAOA \cite{Yu2022}. In contrast, our approach updates the warm-started proposal circuit within an iterative optimization procedure that combines a Metropolis acceptance rule with replica exchange. As such, we offer a novel approach to combinatorial optimization with near-term quantum computers.

\begin{figure*}[ht]
    \centering
    \includegraphics[width=\textwidth]{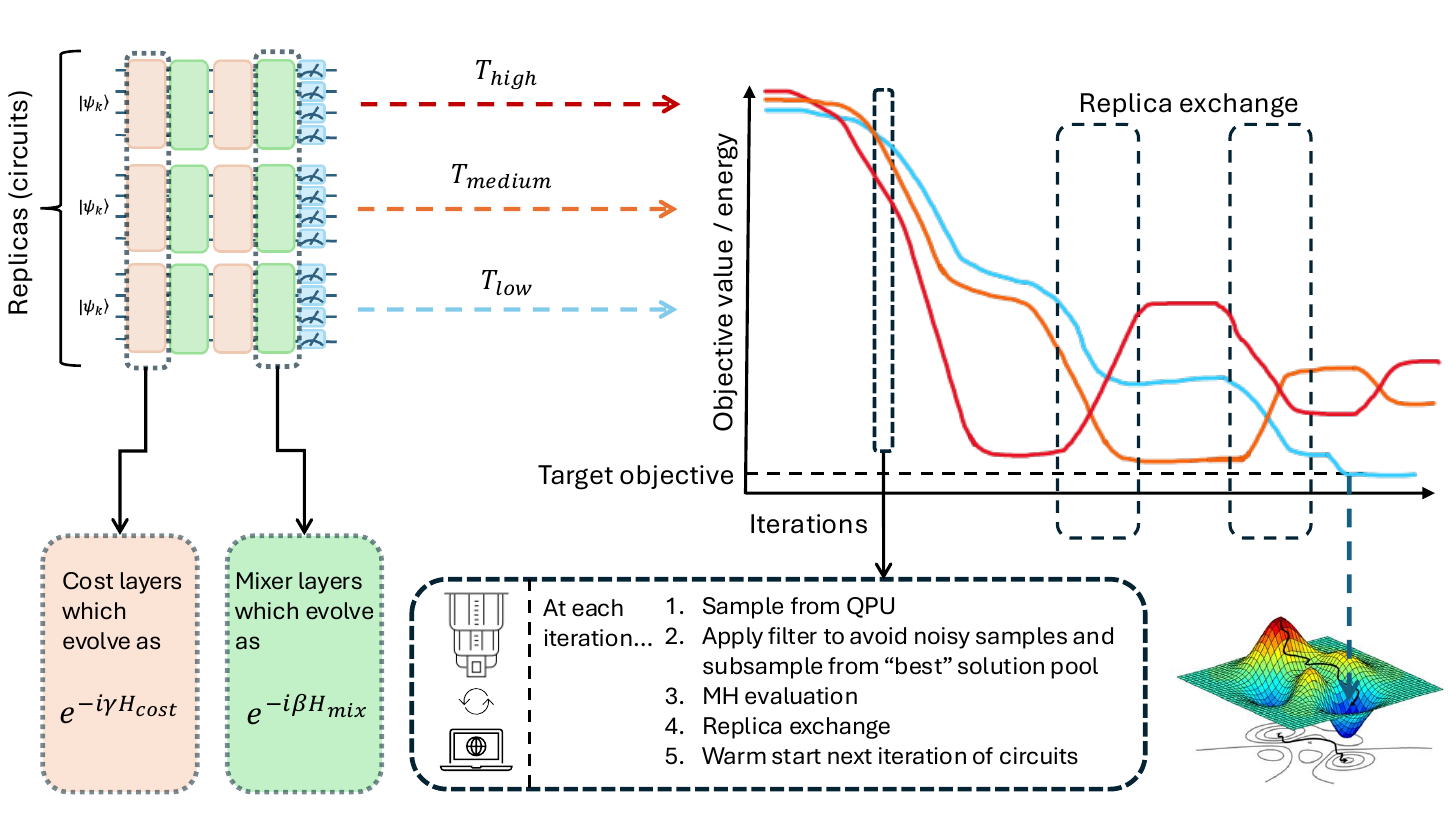}
    \caption{The heuristic quantum optimization workflow. We see a hot, medium and cold replica working together such that the coldest replica eventually converges to the target objective. Objective values of proposed solutions to the MIS problem are plotted against iterations. Each iteration requires drawing samples from the QPU and then classically processing these counts, necessitating a close exchange of information between quantum and classical hardware.}
    \label{fig:workflow}
\end{figure*}

We structure this work as follows. In Section~\ref{sec:background}, we provide background material and present our methods in Section~\ref{sec:methods}. We explain how QeMCMC, warm-starting quantum optimization and parallel tempering work together to form the basis of our workflow in Section~\ref{sec:workflow}, accompanied by Figure~\ref{fig:workflow}. In Section~\ref{sec:hardware_results}, we test the applicability of the algorithm to MIS. We provide comparisons to known classical methods and classical implementations of the algorithm, and discuss
anticipated scaling, or feasibility with increasing problem
size, in Section~\ref{sec:comparison_scaling}. We conclude in Section~\ref{sec:conclusion}. 

\section{Background} \label{sec:background}

\subsection{Markov chain Monte Carlo methods for sampling Boltzmann distributions} \label{sec:MCMC}

Consider a discrete optimization problem defined by an energy or cost function $E(s)$ for configurations $s \in \{-1,+1\}^n$, where the objective is to minimize this energy. Once an energy function is defined, we may associate to it the Gibbs distribution, or equivalently the Boltzmann distribution of a classical Ising model,
\begin{equation}
\mu(s) = \frac{e^{-E(s) / (k_B T)}}{Z},
\label{eq:boltzmann_distribution}
\end{equation}
where $Z = \sum_{s} e^{-E(s) / (k_B T)}$
is the partition function ensuring normalization, and $k_B$ is the Boltzmann constant. At low temperatures, i.e., $T \to 0$, the distribution in Eq.~\eqref{eq:boltzmann_distribution} concentrates its probability mass on configurations that minimize $E(s)$, which provides a connection between probabilistic sampling and optimization. Constraints may be incorporated into the energy landscape, e.g., by adding penalty terms of the form
\begin{equation}
    E(s) = E_0(s) + \lambda\,C(s),
    \label{eq:penalties}
\end{equation}
where $E_0(s)$ denotes the objective, $C(s)$ tracks constraint violations, and $\lambda$ is a sufficiently large penalty factor, or Lagrange multiplier. In the limit $T \to 0$, the Gibbs distribution concentrates probability mass on low-energy configurations, biasing sampling toward high-quality solutions. 

In general, constructing a low-temperature Gibbs state from which to sample is computationally hard. Therefore, in practice, MCMC methods are widely used to sample from Boltzmann distributions~\cite{MCMC_Geman_1984}. MCMC avoids explicitly evaluating the target distribution $\mu$ and its associated partition function, $Z$, since this requires summing over an exponentially large configuration space, and thus, exact computation would require $\Omega(2^n)$ time. Instead, MCMC constructs a Markov chain that evolves through the space of spin configurations by proposing transitions from a current configuration $s$ to a candidate configuration $s'$, according to a fixed transition probability, $P(s' \mid s)$.
A common approach is a two-step proposal and acceptance procedure
\begin{equation}
P(s' \mid s) = Q(s' \mid s)\, A(s' \mid s).
\label{eq:transition_matrix}
\end{equation}
Here, $Q(s' \mid s)$ denotes the conditional probability of proposing candidate transitions from $s$ to $s'$, and $A(s' \mid s)$ is an acceptance ratio. $A$ can be defined in numerous ways, with a popular example being the \textit{Metropolis Hastings (MH)} acceptance probability
\begin{equation}
A(s' \mid s) = \min\!\left( 1, \frac{\mu(s')\, Q(s \mid s')}{\mu(s)\, Q(s' \mid s)} \right).
\label{eq:accept_reject}
\end{equation}
To ensure this procedure yields meaningful samples, the transition probabilities $P$ must be designed such that the chain converges to our target distribution, $\mu$, from Eq.~\eqref{eq:boltzmann_distribution}. This convergence is guaranteed under standard conditions—namely irreducibility and aperiodicity—and by enforcing the detailed balance condition,
\begin{equation}
P(s' \mid s)\, \mu(s) = P(s \mid s')\, \mu(s'),
\label{eq:detailed_balance}
\end{equation}
for all $s \ne s'$ \cite{QeMCMC_Layden_2023}. This detailed balance constrains $P$ such that the Markov chain converges to $\mu$ as its stationary distribution.
In this regime, samples are statistically correct, meaning that they are drawn from the target, $\mu$, and computed averages converge to the true expectation values without systematic bias \cite{QeMCMC_Layden_2023}. 

In the worst case, computation time required to obtain independent samples for the MH variant of MCMC, described above, scales exponentially with system size. Yet, in practice, it is often possible to choose good proposal distributions that exhibit good convergence. This forms the basis of widely-used classical optimization heuristics such as simulated annealing and parallel tempering \cite{MH_1953, MCMC_1970, MCMC_Textbook_2009}. An end-to-end example of  sampling from the Boltzmann distribution with our implementation of QeMCMC is provided in Appendix~\ref{appendix:Boltzmann_sampling}. 

\subsection{Maximum Independent Sets} \label{sec:MIS}

In this manuscript, we study MIS problems, which we will now formulate as a Quadratic Unconstrained Binary Optimization (QUBO) problem that can be mapped to a classical Ising model. This mapping enables us to interpret the problem in terms of an energy landscape and to sample from the corresponding Boltzmann distribution, where low-temperature sampling concentrates on configurations with a minimized energy.

Let $G=(V, E)$ be a graph with vertices $V=\left\{v_1, \ldots, v_n\right\}$ and edges $E \subset V \times V$. 
An independent set of vertices of $G$ is a subset of $V$ such that no pair of vertices in the subset is connected via an edge in $E$. A maximum independent set is an independent set of maximal size.
Now, suppose a binary decision variable $x_i \in \{0,1\}$ for each vertex $v_i$, where $x_i = 1$ if $v_i$ is included in the selected subset of vertices, and $x_i = 0$ otherwise. Then, the MIS problem can be formulated as
\begin{equation}
    \max_{x} \sum_{i=0}^{|V|-1} x_i ,
    \label{eq:independent-set-objective}
\end{equation}
subject to the constraints
\begin{equation}
    x_i x_j = 0 \qquad \forall\, (v_i, v_j) \in E,
    \label{eq:mis_constraint}
\end{equation}
which enforce the selected subset of vertices to be an independent set.
Further, we can formulate this as the following QUBO:
\begin{equation}
    \max_{x} \left( \sum_{v_i \in V} x_i 
    \;-\; \lambda \sum_{(v_i, v_j)\in E} x_i x_j \right),
    \label{eq:lagrangian_form}
\end{equation}
where the first term is the objective and the second term enforces the constraints as long as $\lambda$ is sufficiently large, e.g., $\lambda \geq 2$, i.e., the maximum in Eq.~\eqref{eq:lagrangian_form}, indeed gives an optimal MIS solution.

We convert this QUBO to the Hamiltonian 
\begin{equation}
    \
    H_{\text{cost}} = H_{\text{objective}} + \lambda H_{\text{constraint}},
    \
    \label{eq:H_cost}
\end{equation}
where $H_{\text{objective}}$ and $H_{\text{constraint}}$ are derived and explained using the standard mapping in Appendix~\ref{appendix:MIS_mapping}. 

In this form, each spin configuration corresponds to a candidate solution and has an associated energy $E(s)$, obtained by evaluating the Ising Hamiltonian on that configuration. This energy function defines the Ising model and induces a Boltzmann distribution over configurations, as introduced previously in Eq.~\eqref{eq:boltzmann_distribution}. Sampling from this distribution at temperatures, $T$, allows exploration of the solution space. 

In the limit as $T \to 0$, the distribution concentrates on configurations that minimize $E(s)$, which corresponds to optimal solutions of MIS. Here, we choose values of the Lagrange multiplier, $\lambda$, that do not directly enforce a feasible subspace of configurations, but instead penalize solutions with constraint violations. The choice of this term involves balancing the objective and constraints of our problem, trading feasibility with cost of solutions. The process used to optimize for both objective and constraints is described in Appendix~\ref{appendix:circuit_opt}. 

\section{Methods} \label{sec:methods}

\subsection{Quantum-enhanced Markov chain Monte Carlo} \label{sec:qMCMC}

In Ref.~\cite{QeMCMC_Layden_2023}, the authors propose a quantum-enhanced MCMC scheme in which a quantum circuit is used to generate the proposal distribution. This work was motivated by the fact that classical MCMC approaches struggle to explore rugged energy landscapes at low $T$, which are typical when studying spin glass behavior. The quantum-enhanced MCMC algorithm generates proposed configurations by sampling from a quantum circuit, drawn from a discretized annealing Hamiltonian \cite{QeMCMC_Layden_2023}:
\begin{equation}
    H = (1 - \kappa)\,\alpha\, H_{\text{cost}}
    + \kappa\, H_{\text{mix}} .
    \label{eq:Hamiltonian}
\end{equation}
There is evidence that this quantum sampling procedure can accelerate convergence in spin glasses at low $T$ with up to a quartic advantage over similar classical MCMC methods. Here, $H_{\text{cost}}$ encodes the classical model, $H_{\text{mix}}$ is a mixing Hamiltonian, $\kappa \in [0,1]$ controls the relative weights of both operator terms and $\alpha$ is a normalization factor that rescales the objective term such that its magnitude is comparable to that of the mixer Hamiltonian. More details are found in Ref.~\cite{QeMCMC_Layden_2023}. We can view the penalized energy function introduced in Eq.~(\ref{eq:penalties}) as the classical counterpart of this Hamiltonian.

It is then possible to obtain a new configuration $s'$ by preparing the computational basis state $\ket{s}$ on the quantum device, and applying unitary $U$, given by
\begin{equation}
    U(t) = e^{-i H(\kappa)\, t} .
    \label{eq:evolution_op}
\end{equation}
We then measure the state $U\ket{s}$ to sample the configuration proposal $s'$. Crucially, a key insight made by the authors in Ref.~\cite{QeMCMC_Layden_2023} included that we can avoid explicitly computing the ratio of transition probabilities, $Q$, making the computation of $A$ in Eq.~\eqref{eq:accept_reject} tractable. This finding enabled much of the authors' novel experiments and results. This is achieved by constructing circuits from Eq.~(\ref{eq:Hamiltonian}), where $U$ satisfies the symmetry constraint 
\begin{equation}
|\langle s' | U | s \rangle| = |\langle s | U | s' \rangle|.
\label{eq:symmetry_constraint}
\end{equation} 
The ratio of distributions in Eq.~(\ref{eq:accept_reject}) then cancels out by setting $Q(s' \mid s) = Q(s \mid s')$ for all configurations $s, s' \in \{-1,1\}^n$. The symmetry condition in Eq.~\eqref{eq:symmetry_constraint} ensures reversibility of transitions between configurations, or detailed balance, which is required to guarantee convergence to the target distribution, $\mu$, as defined in Eq.~\eqref{eq:boltzmann_distribution}. Each qubit is measured in the computational basis, denoting the outcome $s'$, before choosing whether to accept or reject the proposed configuration using the acceptance probability defined in Eq.~\eqref{eq:accept_reject}. If $s'$ is rejected, the chain remains at the current state $s$.

We note that the implementation of QeMCMC in our algorithm is heuristic, involving a breakdown of detailed balance, which  impacts the associated acceptance probability defined in Eq.~\eqref{eq:accept_reject} and the overall search behavior. This point is addressed in Appendix~\ref{appendix:heuristic_qemcmc}. 

\subsection{Warm-starting QAOA} \label{sec:warm starting}

Using a first-order Trotter decomposition of the Hamiltonian in Eq.~\eqref{eq:Hamiltonian}, the evolution operator $U$ defined in Eq.~\eqref{eq:evolution_op} can be constructed much like QAOA. Thus, when we apply the unitary to the basis state $\ket{s}$, like $U\ket{s}$, this resembles warm-starting QAOA. Warm-starting QAOA has proved a popular technique in the recent quantum optimization literature \cite{WarmStarting_Egger_2021, WarmStarting_Tate_2022, WarmStarting_Tate_2023, WarmStarting_Cain_2023, WarmStarting_Feeney_2024, WarmStarting_Tate_2024_1, WarmStarting_Tate_2024_2, WarmStarting_Bhattacharyya_2025}. 
The original idea comes from leveraging good solutions determined by a (usually classical) algorithm to initialize another (usually quantum) algorithm, to further improve the results. 
In Ref.~\cite{WarmStarting_Egger_2021}, the authors initialize a QAOA circuit from bitstring $s \in \{0,1\}^n$, either obtained classically, or in our case, from a previously accepted quantum sample. Since the QAOA cost operator has no impact on a state made of a single computational basis state, the authors introduce a regularization parameter $\varepsilon$, which instead prepares a superposition of states in a neighborhood of $s$. We control the spread of the quantum state around $s$ by setting $\varepsilon \in (0,1/2)$ and defining a softened version of $s$, where for each bit $i$ of configuration $s$, we define
\[
\tilde{s}_i =
\begin{cases}
\varepsilon, & s_i = 0, \\
1 - \varepsilon, & s_i = 1.
\end{cases}
\]
The softened solution $\tilde{s}$ is then used to define single-qubit rotation angles,
\[
\theta_i = 2\arcsin\!\bigl(\sqrt{\tilde{s}_i}\bigr),
\]
producing the state
\begin{equation}
\ket{\psi(\theta)}
= \bigotimes_{i=1}^n R_y(\theta_i)\ket{0}
= \bigotimes_{i=1}^n \left(\cos\frac{\theta_i}{2}\,\ket{0} + \sin\frac{\theta_i}{2}\,\ket{1}\right),
\label{eq:warm_start_state_v1}
\end{equation}
or, equivalently,
\begin{equation}
\ket{\psi(\tilde{s})}
= \bigotimes_{i=1}^n \left(\sqrt{1-\tilde{s}_i}\,\ket{0} + \sqrt{\tilde{s}_i}\,\ket{1}\right).
\label{eq:warm_start_state_v2}
\end{equation}
The same $\theta_i$-dependent $R_y$ rotations are used in both the warm-started initial state and the associated mixer. The mixer is adjusted such that its ground state is $\ket{\psi(\tilde{s})}$. This results in solutions to MIS with a bias towards some known good starting point \cite{WarmStarting_Egger_2021}.

\subsection{Parallel tempering} \label{sec:parallel tempering}

When sampling from the Boltzmann distribution of Ising models at low temperatures, the energy landscape typically becomes rugged, characterized by numerous local minima separated by high energy barriers. At low temperatures, classical MCMC methods generally suffer from slow convergence because of long mixing times, i.e., the number of steps a Markov chain must run before its distribution is close enough to the stationary distribution such that samples can be considered representative. In addition, classical MCMC methods may become trapped in local energy minima at low temperatures. To mitigate this behavior, parallel tempering or replica exchange is employed in MCMC procedures \cite{swendsen1986replica, geyer1991interface, hukushima1996exchange}. This approach introduces multiple replicas of the system, each at a different temperature, where the high-temperature replicas explore the energy landscape aggressively, while low-temperature replicas ensure accurate sampling of the target distribution, by closely tracking local minima. Parallel tempering is described in more detail in Refs.~\cite{earl2005parallel, brooks2011handbook}.

To begin, we consider $N$ replicas of the same system, where each replica $i \in \{1, \ldots, N\}$ represents the same Ising model but at a different temperature $T_i$. Temperatures are typically ordered $T_1 < T_2 < \cdots < T_N$. Each replica $i$ evolves independently using its own Markov chain, targeting the Boltzmann distribution, $\mu$, defined in Eq.~\eqref{eq:boltzmann_distribution}. A single iteration of the parallel tempering algorithm then consists of two steps:

\begin{enumerate}
    \item \textit{Local MCMC updates:} Each replica $i$ performs standard MCMC updates at its own temperature.
    \item \textit{Replica exchange stage:} Propose exchange between selected pairs of replicas $i$ and $j$ (see Appendix~\ref{appendix:pt_setup}), which can help low energy replicas escape local minima, with probability 
    \begin{equation}
        \begin{split}
        &\qquad A_{\text{exchange}}(s_i, s_j) =\\
        &\qquad \min\left(1, \exp\left[ (\frac{1}{T_i} - \frac{1}{T_j})\left(E(s_j) - E(s_i)\right) \right] \right).
        \end{split}
        \label{eq:swap_acceptance}
    \end{equation}
\end{enumerate}
This allows a high temperature Markov chain to discover a ``good" candidate solution and transfer it to a low temperature chain, or replica, which will accurately sample around it. These two steps are repeated until the stopping criterion has been satisfied, i.e., we converge to the target objective. 

\subsection{Mitigating noisy samples} \label{sec:noise_mitigation}

For sampling experiments running on noisy quantum computers, there are fewer known techniques for mitigating the impact of noise than for computing expectation values, see Refs.~\cite{EM_Temme_2017, EM_Endo_2018, EM_Kandala_2019, EM_Endo_2021}. Recently, the authors of Ref.~\cite{CVaR_2_Barron_2024} propose to increase the number of samples by $1/\sqrt{\upsilon}$, where $\upsilon$ measures the noise strength in the circuit, and prove bounds on noise free expectation values \cite{CVaR_1_Barkoutsos_2020, CVaR_2_Barron_2024}. The noise strength $\upsilon$ can be measured explicitly or estimated using gate errors provided, e.g., by the \href{https://quantum.cloud.ibm.com}{IBM Quantum Platform} API following compilation of the circuit \cite{McKay_Benchmarking_2023}. By taking $1/\sqrt{\upsilon}$ number of samples, the theory explains that we expect at least one sample from the noise-free distribution corresponding to the target quantum circuit, i.e., in an optimization setting, the additional shots guarantee that we have at least the same performance as in the corresponding noise-free experiment. Where possible, we take the total number of shots or samples suggested via this method in our experiments. Where the requested number of samples is unfeasible, such as when circuits are very deep, we choose to use $10,000$ shots, such that the runtime remains practical. This makes sense for an optimization setting, since not all errors affect the final solution, and noise can help to move closer to the target objective in some cases. Thus, using fewer shots than this theoretical quoted bound can still be successful.

We evaluate and rank the objective values of all samples and choose to keep the lowest-energy, i.e., the best, samples. This introduces a bias towards good solutions, which is desirable in the context of our optimization task. We test both taking the best single low-energy sample from the distribution of results, as well as subsampling from a portion of, say $10$, of the lowest-energy bitstrings, to maintain some balance of exploration of the cost landscape, i.e., escaping local minima, and exploitation of good solutions. Duplicate solutions at the cutoff point are retained, such that we still favor resampled solutions from the circuit.

\begin{table*}[t]
\centering
\resizebox{\textwidth}{!}{%
\begin{tabular}{|l|c|c|}
\hline
\textbf{Parameter} &
\textbf{Symbol} &
\textbf{Optimization method} \\
\hline

Cost Hamiltonian evolution angle (QAOA) & $\gamma$ &
Classical optimization pipeline \cite{qaoa_training_pipeline} \\

Mixer Hamiltonian evolution angle (QAOA) & $\beta$ &
Classical optimization pipeline \cite{qaoa_training_pipeline}  \\

MIS constraint term (Lagrange multiplier) & $\lambda$ &
Classical optimization pipeline \cite{mis_benchmarking_pipeline} \\

Regularization parameter for warm-starting & $\varepsilon$ &
Experimentally informed, see Appendix~\ref{appendix:circuit_opt} \\

Temperature range used for parallel tempering setup & $[T_\text{low},  T_\text{high}]$ &
Determined via formulae, see Appendix~\ref{appendix:pt_setup} \\

Number of replicas used for parallel tempering setup & $N_{\text{replicas}}$ &
Design choice, see Appendix~\ref{appendix:pt_setup} \\

Replica exchange regularity & $\Delta t_{\text{swap}}$ &
Experimentally informed, see Appendix~\ref{appendix:pt_setup} \\
\hline
\end{tabular}
}
\caption{Summary of parameters, their roles and optimization method for each problem instance.}
\label{table:parameters}
\end{table*}

\subsection{Circuit construction} \label{sec:circuit_construction}

We construct circuits from Eq.~\eqref{eq:Hamiltonian} that resemble QAOA, using a first-order Trotter decomposition, since Trotter error is not relevant. Since 
\(
[H_{\text{cost}}, H_{\text{mix}}] \neq 0,
\)
Eq.~\eqref{eq:Hamiltonian}.
cannot be implemented exactly as a single exponential. The Lie–Trotter decomposition,
\begin{equation}
e^{-i t H}
\approx
e^{-i t (1-\kappa)\alpha H_{\text{cost}}}\,
e^{-i t \kappa H_{\text{mix}}}
\;+\; \mathcal{O}(t^2),
\label{eq:trotter_1st_order}
\end{equation}
divided into two identical Trotter steps yields
\begin{equation}
e^{-i t H}
\approx
\left(
e^{-i \gamma H_{\text{cost}}}\,
e^{-i \beta H_{\text{mix}}}
\right)^2,
\label{eq:p2_qaoa_trotter}
\end{equation}
with fixed angles
\begin{equation}
\gamma = \tfrac{t}{2}(1-\kappa)\alpha,
\qquad
\beta = \tfrac{t}{2}\kappa .
\end{equation}
This construction has the form of a \(p=2\) QAOA circuit with parameters,
\(
\gamma_1=\gamma_2\equiv\gamma
\)
and
\(
\beta_1=\beta_2\equiv\beta.
\)
Details on how $\gamma$ and $\beta$ parameters are trained are found in Appendix~\ref{appendix:circuit_opt}. Our unitary in Eq.~\eqref{eq:evolution_op} has the form 
\begin{equation}
U_{\text{QAOA}}
=
e^{-i\beta_2 H_{\text{mix}}}e^{-i\gamma_2 H_{\text{cost}}}
e^{-i\beta_1 H_{\text{mix}}}e^{-i\gamma_1 H_{\text{cost}}}.
\label{eq:qaoa}
\end{equation}
Throughout this work, we use a warm-started variant of the standard transverse-field mixer, $H_{\text{mix}} = \sum_{i=1}^n X_i$. $H_{\text{cost}}$ is defined in Eq.~\eqref{eq:H_cost}.

Since our circuits structurally resemble QAOA, in order to run QeMCMC-like experiments, we cannot simply initialize $U_\text{QAOA}$ in the computational basis state $\ket{s}$, as per Ref.~\cite{QeMCMC_Layden_2023}. When using this initialization with the form of the standard transverse-field mixer, we observe limited exploration of the solution space as noted in Ref.~\cite{WarmStarting_Cain_2023}, motivating the warm-starting procedure described herein. Instead, we adapt the warm-starting QAOA procedure from Section~\ref{sec:warm starting}, restricting the sample space used for our quantum proposal distribution, $Q$, to the best, low-energy portion. This, in combination with the bias introduced from Section~\ref{sec:noise_mitigation}, distinguishes us from the original goals of QeMCMC, developed in the context of sampling from Boltzmann distributions. Instead, we seek the ground state of an Ising Hamiltonian framing our optimization problem. We propose candidate updates $s_k \to s_{k+1}$, according to our proposal distribution $Q(s_{k+1} \mid s_k )$. This is done by applying the unitary, $U_\text{QAOA}$, in Eq.~\eqref{eq:qaoa} to $\ket{s_k}$, and defining transition probabilities from the resulting measurement statistics. Crucially, since we introduce these asymmetric proposal distributions in our sampling procedure, the convergence guarantees of QeMCMC originally positioned by the authors in Ref.~\cite{QeMCMC_Layden_2023} no longer hold. Our approach is, therefore, purely a heuristic one. 

Finally, as MIS problem instance sizes grow, the width and depth of our QAOA ansaetz representing the full graph become too large to run on quantum hardware. Therefore, we construct ansaetz from a simplified cost Hamiltonian obtained by dropping Pauli terms that require too many SWAP gates to implement. For further details, see Appendix~\ref{appendix:ansaetze}. 

\subsection{Experimental Workflow} \label{sec:workflow}

We combine these methods to build the algorithm visualized in Figure~\ref{fig:workflow} and summarized herein. Firstly, we define a number of identical replica circuits and tune the temperature ladder for these circuits to operate, as described in Section~\ref{sec:parallel tempering} and Appendix~\ref{appendix:pt_setup}. For each circuit, we sample an initial configuration uniformly at random, i.e., $s_0 \sim \mathcal{U}(\{-1,+1\}^n)$, and then proceed as follows:

\begin{enumerate}
    \item We sample from our parametrized replica circuits derived from Eq.~\eqref{eq:Hamiltonian}. Parameters are classically pre-trained, as described in Appendix~\ref{appendix:circuit_opt}.
    \item We take many samples, in order to mitigate the impact of noise on the quantum device, as per the theory described in Section~\ref{sec:noise_mitigation}. Then, we evaluate the energy $E(s)$ for each sample, filter for a certain number of low energy samples, and randomly subsample one configuration from these. 
    \item Next, we classically evaluate the proposed solution using the MH criteria defined in Eq.~\eqref{eq:accept_reject}.
    \item At regular intervals, replicas are allowed to exchange accepted solutions according to the probability defined in Eq.~\eqref{eq:swap_acceptance}, which helps low temperature replicas escape local minima.
    \item The accepted solution $s_k$ is used to generate subsequent candidate updates $s_k \to s_{k+1}$, according to the proposal distribution $Q(s_{k+1} \mid s_k )$, as described in Section~\ref{sec:MCMC}. This is done using the warm-start QAOA procedure explained in Section~\ref{sec:warm starting}.
\end{enumerate}

We repeat for a maximum of $200,000$ iterations, or until convergence. Tuning algorithm parameters for each problem instance strongly influences mixing efficiency and convergence \cite{earl2005parallel,kone2005selection,nadler2008optimized,chodera2011replica}. We use analytical guidelines drawn from the literature, as well as heuristic tuning. A summary of these parameters is presented in Table~\ref{table:parameters}, and further details regarding their optimization can be found in Appendices \ref{appendix:circuit_opt} and \ref{appendix:pt_setup}. 


\section{Results} \label{sec:results_intro}

We study five different problem instances sourced from the \textit{Network Repository}, a large public collection of graph datasets used in machine learning, network science, and graph algorithm benchmarking \cite{network_repository}. 
These graphs are derived from real-world animal social interaction networks, with qubit or node counts $n \in \{17, 39, 52, 80, 117\}$ and corresponding edge counts $m \in \{91, 245, 454, 714, 1028\}$. The graphs share a common provenance and representation as simple, undirected interaction networks. As social networks, they exhibit heterogeneous degree distributions and nontrivial interaction structure, while varying in size and density. 
Together, these instances provide a natural progression in problem size for probing the scalability of our algorithm. 
Many of these problem instances are also included in the QOBLIB \cite{QOBLIB_Koch_2025}.
Preparation and algorithm execution for each problem instance used when testing the heuristic quantum optimization algorithm was costly in both computational resources and time. For this reason, only five problem instances were prepared and tested within the scope of this work. Information regarding the runtime data for these experiments can be found in Appendix~\ref{appendix:runtime_data}. 

We focus our analysis primarily on the $117$-node problem instance, which consists of a fully connected component of $110$ nodes and $7$ isolated nodes. While the isolated nodes could be removed, we keep the full graph without any pre-processing. 
We explore these problem instances using classical replica-exchange MCMC, classical matrix product state (MPS) simulation of the QeMCMC algorithm, and executions on real IBM quantum hardware. We first present classical baseline results using a classical proposal distribution, which then serve as a benchmark for our quantum-enhanced approach. Then, we run our algorithm using both classical MPS simulation and real quantum hardware. Experimental details such as number of replicas, tuning of the temperature ladder (described in Appendix~\ref{appendix:pt_setup}) and regularity of replica exchanges are kept consistent across setups.
For classical MCMC and MPS simulations, we compare both single-shot and multi-shot settings per iteration. 
In contrast, for real quantum hardware, we only consider multiple shots per iteration to account for noise, as motivated in Section~\ref{sec:noise_mitigation}. 
The globally optimal objective values for all problem instances were computed via CPLEX, an advanced classical optimization suite capable of solving the described problem instances to optimality \cite{CPLEX}.

\subsection{Classical baseline results}  \label{sec:classical_algo_results}

The most natural comparison for our heuristic QeMCMC algorithm, is to first understand the capabilities of classical replica exchange MCMC. In this setting, the proposal distribution is constructed via classical bitflips. Various proposal strategies for the classical MCMC setup were explored, including uniformly sampled bitflip proposals of up to 25 bitflips per sample for each replica. These proposals can generate large Hamming-distance moves but are not energy-aware, unlike cluster or isoenergetic moves designed to produce large changes in configuration with relatively small changes in energy. Testing other non-uniform proposal strategies is left out of scope for this work. The most competitive classical configuration was uniform sampling with up to $5$ bitflips per iteration, for which results are presented here. We test the classical setup with both $1$ shot and $10,000$ shots per iteration. In the case where $10,000$ shots are taken per iteration, we randomly subsample from the best ranked ten proposals.

Just as is done for the quantum-enhanced algorithm setting, samples for the first iteration for each experiment are taken from the non-warm started $p=2$ QAOA circuit. This experiment is repeated $10$ times, with $200,000$ iterations per run. 

\begin{table}[h]
\centering
\resizebox{\columnwidth}{!}{%
\begin{tabular}{|l|c|c|c|}
\hline
\textbf{Replica} &
\textbf{Temperature} &
\begin{tabular}{c}
\textbf{Median num.~iterations} \\
\textbf{until convergence} \\
\textbf{(1 shot per iteration)}
\end{tabular} &
\begin{tabular}{c}
\textbf{Median num.~iterations} \\
\textbf{until convergence} \\
\textbf{($10k$ shots per iteration)}
\end{tabular}
\textbf{ } \\
\hline
1 & 0.01 & 6293 & 5101 \\
2 & 0.11 & 7554 & 5051 \\
3 & 0.21 & 8191 & 6447 \\
4 & 0.51 & 8576 & 7006 \\
5 & 1.01 & --- & 9196 \\
\hline
\end{tabular}
} 
\caption{Replica convergence statistics for 5 replicas across 10 independent repeats of the classical replica exchange MCMC experiment for 117 node graph instance, where either $1$ shot or $10,000$ shots are taken per iteration of the algorithm, with random subsampling from the best ten proposals.}
\label{table:convergence_117_node_classical}
\end{table}

The lowest number of samples required to converge to the global optimum in the $1$-shot per iteration setting is $6,293$ samples or iterations. With $10,000$ shots per iteration, the lowest number of iterations required to converge to the global optimum is $5,051$ iterations. The overall results for both settings and all temperatures are shown in Table~\ref{table:convergence_117_node_classical}.
While these numbers are close in terms of iterations, the total number of shots is significantly larger in the multi-shot setting. Thus, the additional shots seem to create little additional value in this noise-free setting.


\subsection{Simulation of quantum hardware}  \label{sec:sim_results}

Simulations of the quantum-enhanced algorithm, with quantum circuits composing the proposal distribution, were performed using the \textit{Qiskit Aer} MPS simulator. MPS simulations are subjected to a truncation threshold, which specifies the tolerance for discarding Schmidt coefficients: coefficients are truncated when the sum of the squares of the discarded values falls below the specified threshold \cite{qiskit2024}. Truncation thresholds ranged from $10^{-6}$ to $10^{-10}$ across all problem instances studied, with a threshold of $10^{-6}$ applied to simulations of the $117$ node problem, for which results are presented below. This threshold was chosen to balance minimizing truncation error against feasible runtimes for the simulations. 
The quantum circuits implemented via MPS are the same ansatz circuits constructed for the hardware experiments, for which the process is described in Appendix~\ref{appendix:ansaetze}. Again, we test both 1-shot setting and 10,000-shot setting with subsampling from the best ten proposals. 

Circuits are initialized in the same way, with initial samples drawn from the non-warm started $p=2$ QAOA circuit. These experiments are repeated $10$ times, with up to $200,000$ iterations.

\begin{table}[h]
\centering
\resizebox{\columnwidth}{!}{%
\begin{tabular}{|l|c|c|c|}
\hline
\textbf{Replica} &
\textbf{Temperature} &
\begin{tabular}{c}
\textbf{Median num.~iterations} \\
\textbf{until convergence} \\
\textbf{(1 shot per iteration)}
\end{tabular} &
\begin{tabular}{c}
\textbf{Median num.~iterations} \\
\textbf{until convergence} \\
\textbf{($10k$ shots per iteration)}
\end{tabular}
\textbf{ } \\
\hline
1 & 0.01 & 91668 & 863 \\
2 & 0.11 & --- & 796 \\
3 & 0.21 & --- & 706 \\
4 & 0.51 & --- & 619 \\
5 & 1.01 & --- & --- \\
\hline
\end{tabular}
} 
\caption{Replica convergence statistics for 5 replicas across 10 independent repeats of the simulated quantum experiment for 117 node graph instance, where either $1$ shot or $10,000$ shots are taken per iteration of the algorithm, with random subsampling from the best ten proposals.}
\label{table:convergence_117_node_sim}
\end{table}

In Table~\ref{table:convergence_117_node_sim}, we see the number of iterations needed to converge to the global optimum for both $1$ shot per iteration and $10,000$ shots per iteration. For $1$ shot per iteration, only the coldest replica managed to converge to the optimal solution after $91,668$ samples or iterations. For $10,000$ shots per iteration, we see that colder replicas have a higher rate of success in converging to their target values, but that a warmer replica, Replica $4$, was able to converge in the fewest number of iterations in total, likely due to its broader search behavior. These results are also visualized in Appendix~\ref{appendix:pt_setup}.
It is interesting that taking multiple samples per iteration with subsampling now strongly affects the number of required iterations to converge, unlike in Section~\ref{sec:classical_algo_results}. We expect this stems from the fact that QAOA-based proposal distributions, while introducing a bias towards good solutions, also produce a tail of much worse ones. Subsampling effectively isolates the high-quality tail of the proposal distribution, making it beneficial beyond merely mitigating noise.


\subsection{Quantum hardware results}  \label{sec:hardware_results}

Finally, we present the results of our heuristic QeMCMC optimization algorithm using IBM quantum hardware. In this setting, we do not explore the $1$-shot regime due to the presence of noise and because it would lead to prohibitively long runtimes. Including circuit preparation overheads on the quantum hardware, taking $1$ shot per circuit takes almost as long as taking $10,000$ shots per circuit. 
Thus, we execute experiments taking $10,000$ shots per iteration followed by the subsampling procedure described in Sec.~\ref{sec:noise_mitigation}. 

\begin{table*}[t]
\centering
\resizebox{\textwidth}{!}{%
\begin{tabular}{|l|c|c|c|c|c|c|c|c|c|}
\hline
\textbf{Device} &
\begin{tabular}{c}
\textbf{Problem} \\
\textbf{instance} \\
\textbf{(nodes)}
\end{tabular} &
\begin{tabular}{c}
\textbf{Num.} \\
\textbf{SWAP} \\
\textbf{layers}
\end{tabular} &
\begin{tabular}{c}
\textbf{2q gate } \\
\textbf{count}
\end{tabular} &
\begin{tabular}{c}
\textbf{2q gate } \\
\textbf{depth}
\end{tabular} &
\begin{tabular}{c}
\textbf{Median num.} \\
\textbf{iterations} \\
\textbf{to converge}
\end{tabular} &
\begin{tabular}{c}
\textbf{Est. shots} \\
\textbf{required} \\
\textbf{per iter}
\end{tabular} &
\begin{tabular}{c}
\textbf{Shots used} \\
\textbf{per iter}
\end{tabular} &
\begin{tabular}{c}
\textbf{Num. cost} \\
\textbf{function} \\
\textbf{evaluations}
\end{tabular} &
\begin{tabular}{c}
\textbf{Post-processing } \\
\textbf{comments}
\end{tabular} \\
\hline

ibm\_aachen & 17 & $14 \text{*}$
& $740$ &
$96$ & 32
& 5 & 5 & $160$
& Best sample taken \\

ibm\_aachen & 17 & $14 \text{*}$
& $740$ &
$96$ & 2
& $5$ & $10,000$ & $20,000$
& Sample from best ten\\

ibm\_boston & 39 & 32
& $3616$ &
$198$ & 12
& $2,065$  & $10,000$ & $120,000$
& Sample from best ten\\

ibm\_boston & 52 & 13
& $1964$ &
$84$ & 22
& $24,503$  & $10,000$ & $220,000$
& Sample from best ten\\

ibm\_boston & 52 & 37
& $5532$ &
$228$ & 37
& $4,280,674$ & $100,000$ & $3,700,000$
& Sample from best ten\\

ibm\_boston & 80 & 13
& $3250$ &
$84$ & 106
& $20,504$ & $10,000$ & $1,060,000$
& Sample from best ten\\

ibm\_boston & 117 & 6
& $1574$ &
$42$ & 151
& $30,426$ & $10,000$ & $1,510,000$
& Sample from best ten\\

ibm\_pittsburgh & 117 & 6
& $1574$ &
$42$ & 114
& $1,034,484$ & $10,000$ & $1,140,000$
& Best sample taken \\

\hline
\end{tabular}
}
\caption{Summary of notable experimental results. All experiments used $5$ replicas. At least three converged repeats were run for each experiment on the hardware devices. Estimated numbers of shots required per iteration are derived from the logic explained in Section~\ref{sec:noise_mitigation}, where differences in how devices experience noise lead to different values for estimated numbers of shots required per iteration. At the time of writing, error rates were lowest for the \textit{Heron r1} devices, followed by \textit{Heron r2} and \textit{Heron r3} generation devices. \newline \text{*} This is the maximum number of SWAP layers for this graph i.e. the full graph is implemented for this instance. }
\label{table:results}
\end{table*}

\begin{figure*}[ht]
    \centering
    \begin{subfigure}[t]{\textwidth}
        \centering
        \includegraphics[width=\linewidth]{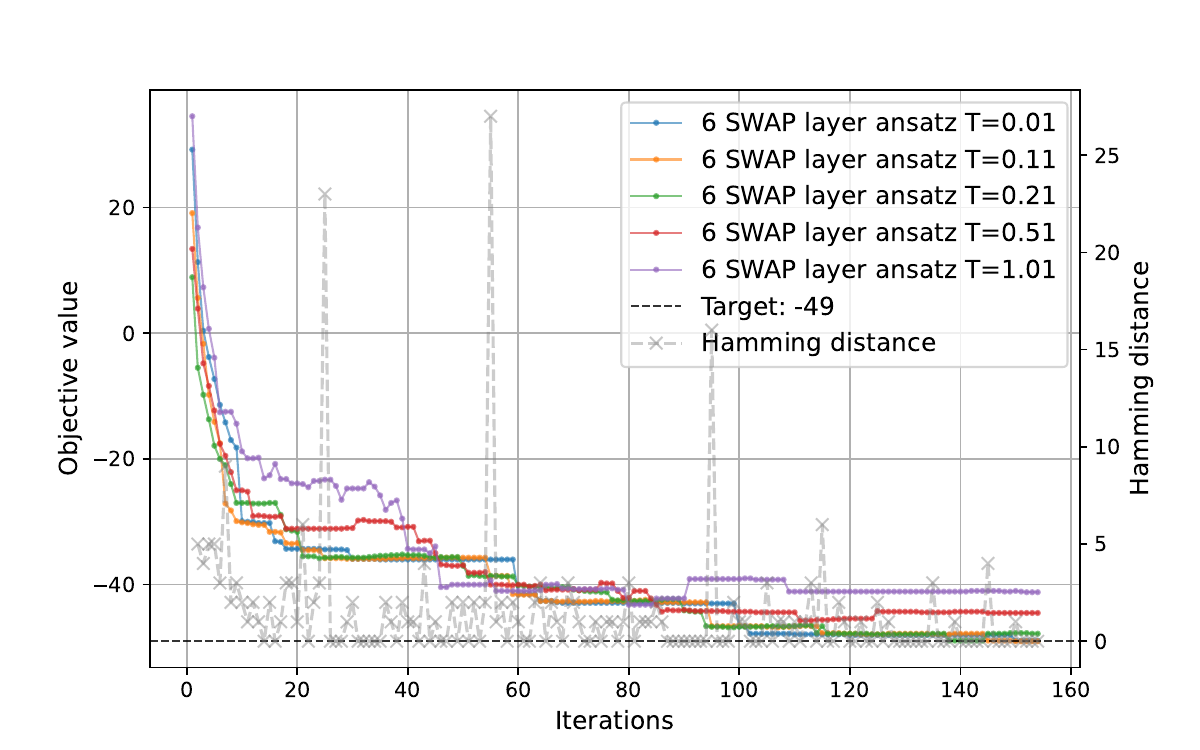}
    \end{subfigure}

    \caption{Parallel tempering experiment run on \texttt{ibm\_boston} for the 117-node MIS problem. The left axis indicates the progress of objective values per iteration. We see the global optimum is found after $151$ iterations, with $10,000$ shots per iteration and a random sample taken from a smaller portion of these. The Hamming distance between accepted solutions is also plotted for each iteration for the winning replica in orange, corresponding to the right axis.}
    \label{fig:hardware_117_nodes}
\end{figure*} 

Results for the different problem instances are shown in Table~\ref{table:results}. These include the maximum number of SWAP layers allowed for the simplified ansaetz for these problems, alongside the associated 2-qubit gate counts and depths of the circuits. We report the total shots taken alongside the number of samples kept for subsampling for different instances and quantum devices, together with their convergence analysis. 
In both sets of experiments using the $117$ node graph, $10,000$ shots were taken on the device, with median convergence counts of $151$ and $114$ iterations, respectively. 
The number of shots taken is significantly less than the estimated number required to compensate for noise. However, since these estimations are upper bounds (see Section~\ref{sec:noise_mitigation}), in practice, we choose $10,000$ shots such that we can expect a reasonable runtime on the device. 


Results for the 117-node problem instance  on \texttt{ibm\_boston} are visualized in Figure~\ref{fig:hardware_117_nodes}. The winning replica that found the optimal solution first is represented in orange. For this replica, the Hamming distance between accepted solutions at each iteration is also displayed. From the data shown in Table~\ref{table:results}, we see that the quantum hardware exceeds expectations in terms of algorithm performance indicated from our classical MPS simulations reported in Table~\ref{table:convergence_117_node_sim}, since convergence was achieved in $151$ iterations, which is far fewer iterations on hardware than $796$ iterations predicted for the same replica via simulation. One possible explanation is that truncation error in this regime has a greater impact on convergence than hardware noise, since we also observed worsening performance with increased truncation error introduced to the MPS experiments via controlled testing.

\subsection{Comparison and scaling analysis} \label{sec:comparison_scaling}

\begin{figure*}[p]
    \centering


    \begin{subfigure}{0.7\textwidth}
        \centering
        \includegraphics[width=\textwidth]{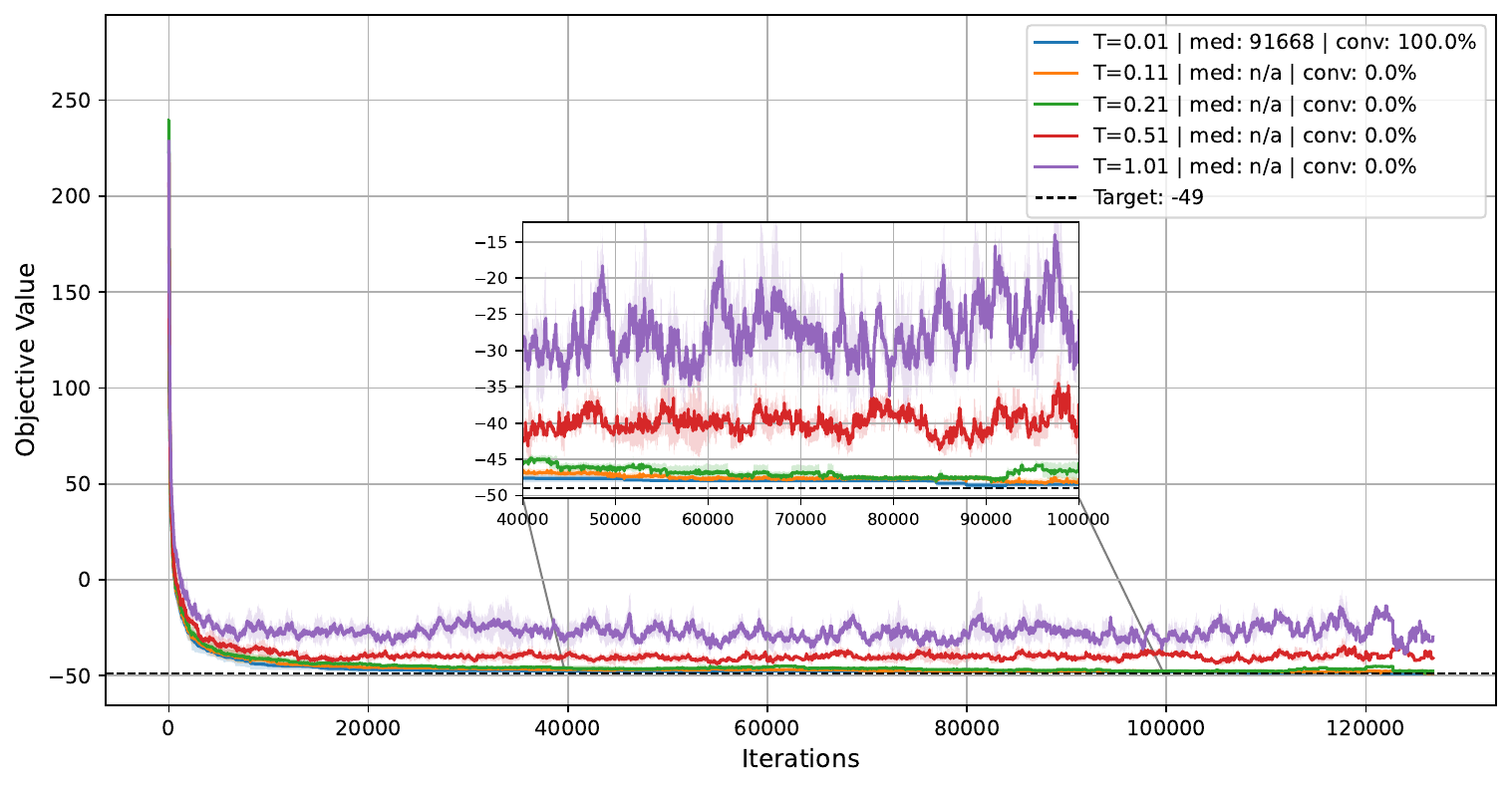}
        \label{fig:117_node_result_1_shot_QeMCMC_simulation}
    \end{subfigure}

    \vspace{0.2em}
    
    \begin{subfigure}{0.7\textwidth}
        \centering
        \includegraphics[width=\textwidth]{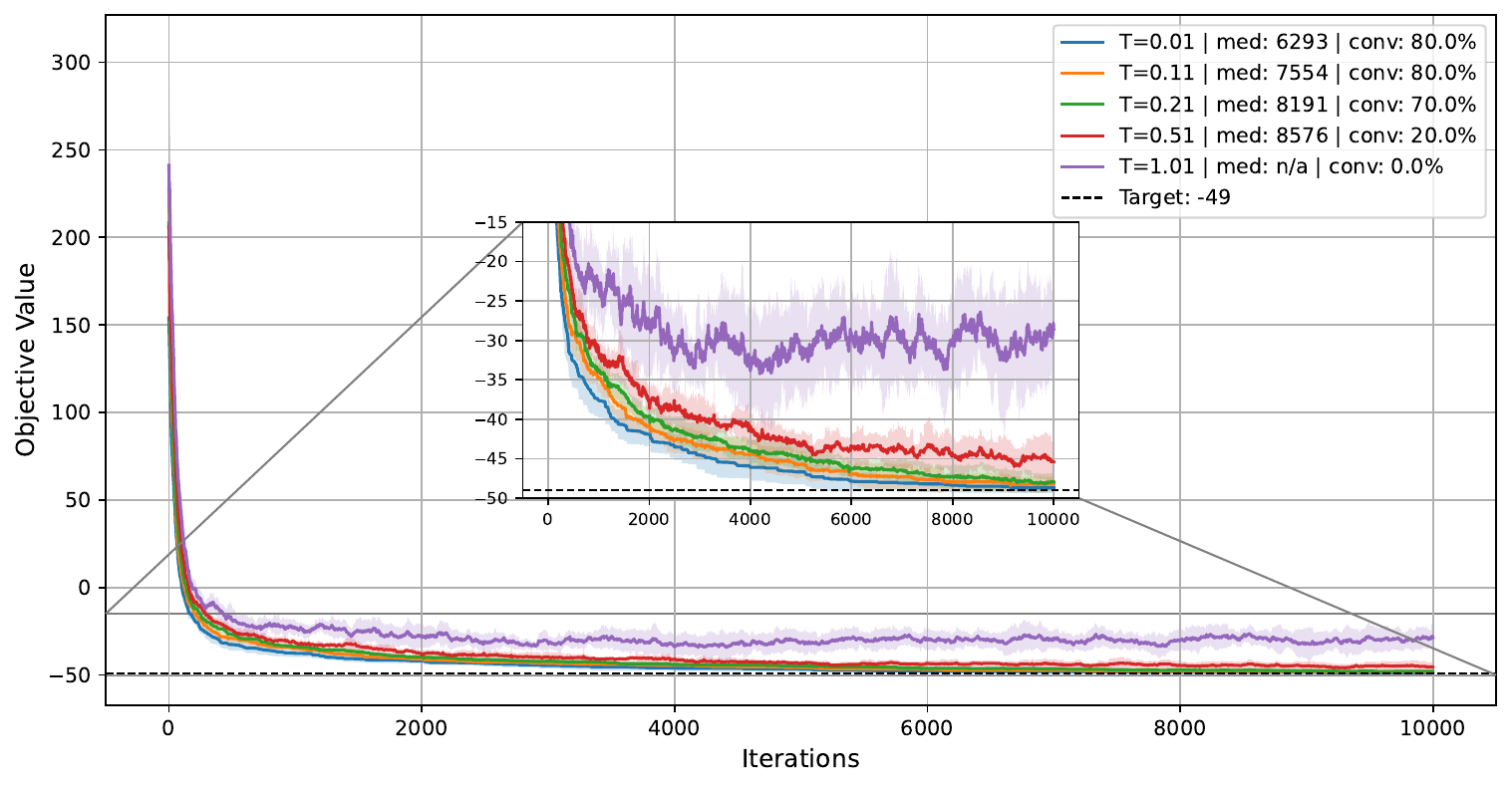}
        \label{fig:117_node_result_1_shot_CMCMC_simulation}
    \end{subfigure}

    \vspace{0.2em}

    \begin{subfigure}{0.7\textwidth}
        \centering
        \includegraphics[width=\textwidth]{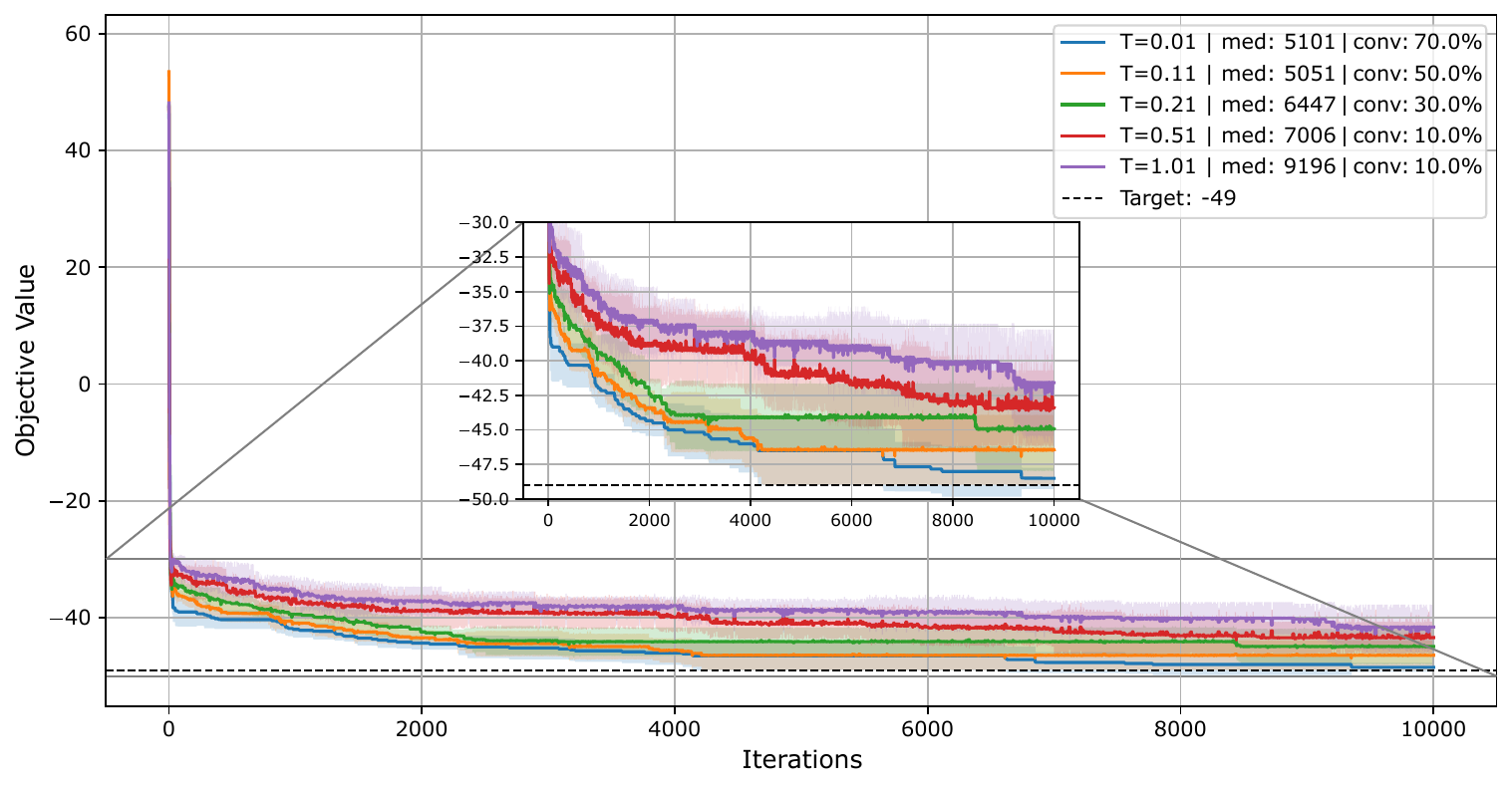}
        \label{fig:CMCMC_117_nodes_hardware_setup_comparison}
    \end{subfigure}


    \caption{Comparison of experimental setups for 117 node problem instance experiments. Simulations of the quantum-enhanced algorithm are compared to classical MCMC methods used to find the target objective. In the top panel, we present an MPS simulation of the quantum-enhanced algorithm. Across ten repeats, the target objective is found after a median of $\sim90,000$ iterations for the coldest replica, with $1$ shot per iteration. The middle panel shows a comparison to a classical MCMC simulation, which also uses $1$ shot per iteration. Across ten repeats, the target objective is found in $4,000 - 8,000$ iterations for the coldest replica, with a median of $\sim6,000$ shots, or iterations. Finally, in the bottom panel, we display another classical MCMC simulation, where across ten repeats, the target objective is found in $4,000 - 8,000$ iterations for the coldest replica, with a median of $\sim5,000$ iterations, in this case using $10,000$ shots per iteration and a random sample taken from the best ten proposals.}
    \label{fig:117_node_results}
\end{figure*}

In this Section, we compare results of the different algorithms introduced herein. There exist various possible metrics to compare algorithm performance across experiments and analyze scaling behavior for increasing problem size, including total wall-clock time for the experiments, or comparing total numbers of shots, samples, cost-function evaluations or algorithm iterations. Some of these statistics are reported in Appendix~\ref{appendix:runtime_data}. 

\begin{figure*}[t]
    \centering
    \begin{subfigure}[t]{0.49\textwidth}
        \centering
        \includegraphics[width=\linewidth]{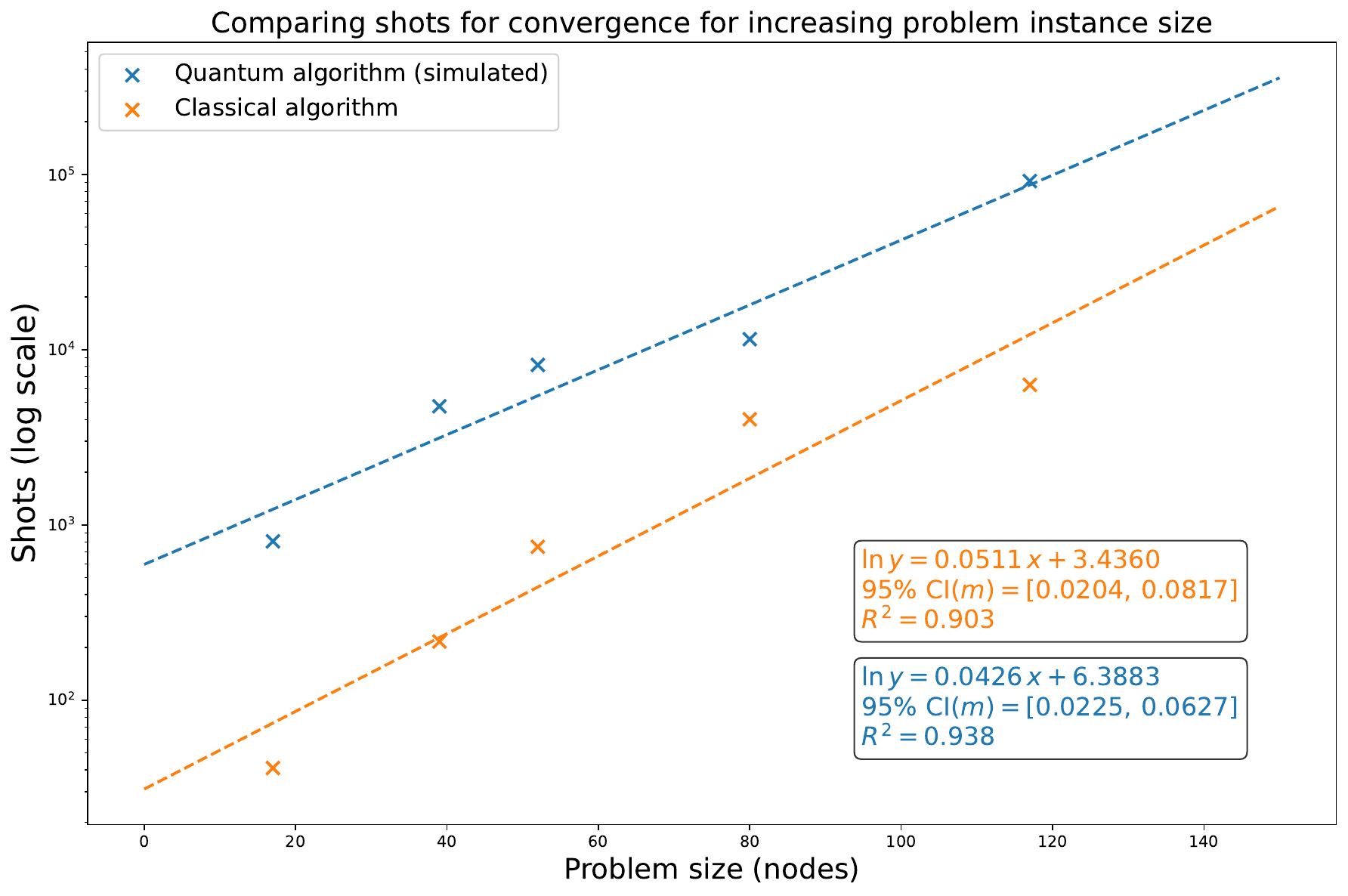}
        \label{fig:scaling_shots}
    \end{subfigure}
    \begin{subfigure}[t]{0.49\textwidth}
        \centering
        \includegraphics[width=\linewidth]{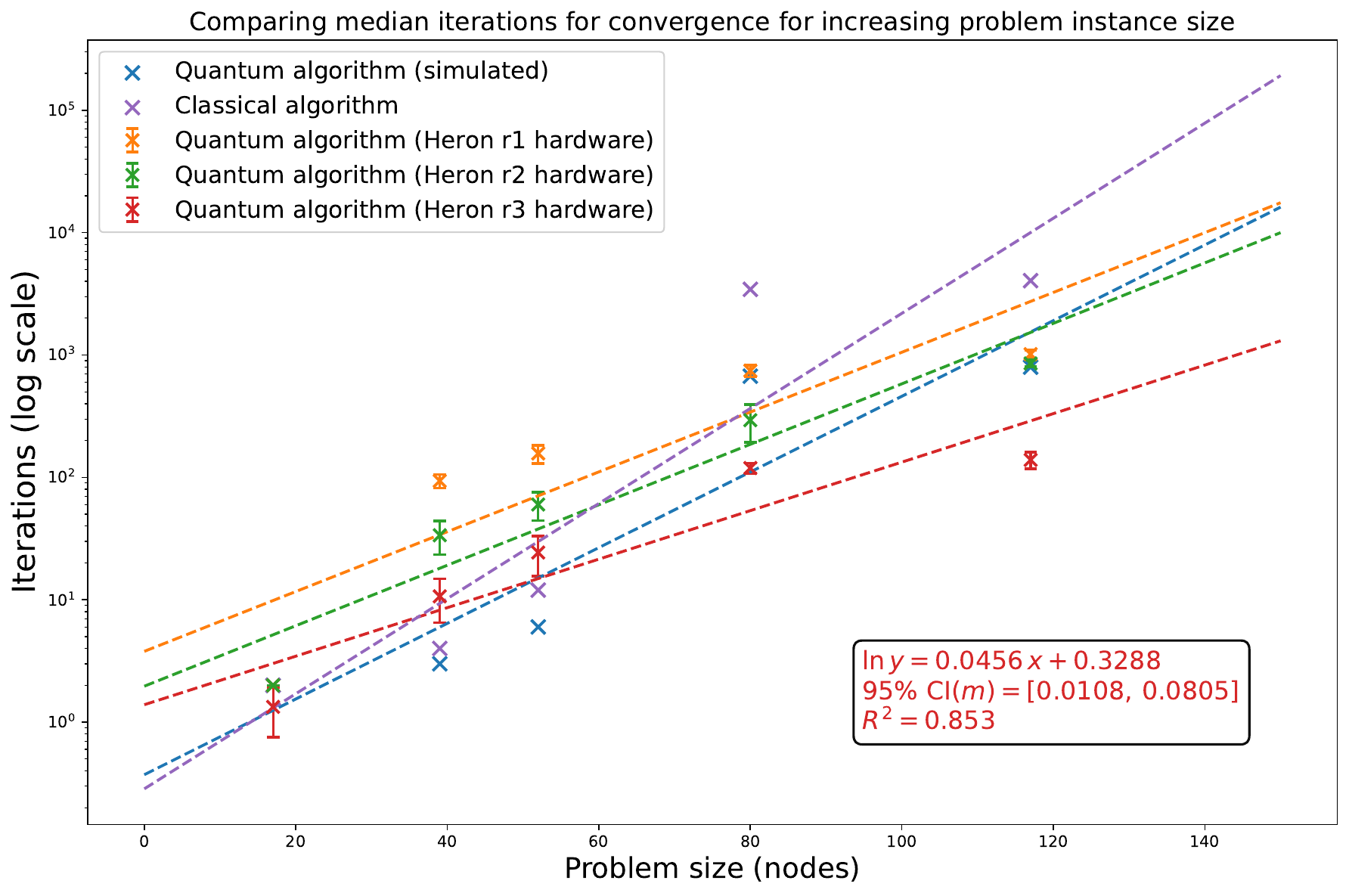}
        \label{fig:scaling_iterations}
    \end{subfigure}

    \caption{A summary of anticipated performance scaling in terms of shots and iterations of the classical vs. quantum-enhanced algorithm. Median iterations to converge are reported for the 17, 39, 52, 80 and 117 node graphs after ten repeats of the classical and quantum-enhanced algorithms, and median iteration counts for quantum hardware are computed using three successful, converged runs only. A line of best fit provides us with a naïve indication of scaling behavior as to how we anticipate each method to perform with increasing problem instance size. On the left, we present anticipated performance scaling in terms of shots for the classical vs. simulated quantum-enhanced algorithm, using the \textit{Qiskit Aer} MPS simulator. On the right, we see anticipated performance scaling in terms of iterations of the classical vs. quantum-enhanced algorithm, both simulated and run on real \textit{Heron r1}, \textit{Heron r2} and \textit{Heron r3} IBM quantum hardware.}
    \label{fig:scaling_summary}
\end{figure*}

In Figure~\ref{fig:117_node_results}, we compare results for the 117-node problem. The top panel shows results for an MPS simulation of the quantum algorithm with $1$ shot per iteration. The experiment converges in a median of $\sim90,000$ samples (or iterations) across ten trials, as can be seen in Table~\ref{table:convergence_117_node_sim}.
This is easily outperformed in $\sim6,000$ samples (or iterations) required by the standard $1$-shot classical MCMC shown in the middle panel of Figure~\ref{fig:117_node_results} and summarized in Table~\ref{table:convergence_117_node_classical}. Finally, we take $10,000$ samples per iteration and subsample from the best ten again for the classical MCMC implementation. The results for this setting are visible in the bottom panel of Figure~\ref{fig:117_node_results}. Although convergence is comparable to the standard $1$-shot setup, the best reported result of $\sim5,000$ iterations is an order of magnitude larger than the equivalent setting for the quantum algorithm, both simulated (see Table~\ref{table:convergence_117_node_sim}) and on device (see Table~\ref{table:results}). Thus, we see that classical approach performs well with $1$-shot per iteration, whereas the quantum algorithm performs better with many shots. Next, we observe how the algorithms scale in terms of iterations and samples with increasing problem size. 
For the classical and simulated quantum algorithms, ten independent trials were run and various statistics were computed for each of the problem instances, as shown in Figure~\ref{fig:scaling_summary}. Median values, equations of the fitted lines, and the quality of these fits are plotted for both number of samples and/or iterations as an indication of exponential growth rates for the five problem instances studied. 

On the left of Figure~\ref{fig:scaling_summary}, in the $1$-shot setting, the classical algorithm outperforms the simulated quantum experiment for each problem instance. However, the fitted exponential growth rate for the MPS simulations is $\sim83\%$ of the classical results, suggesting an early indication of favorable scaling behavior of the quantum algorithm over similar MCMC-type heuristics for the studied problems. However, the 95\% confidence intervals for the fitted slopes overlap, so the observed difference does not establish a statistically significant scaling advantage.

In the right panel of Figure~\ref{fig:scaling_summary}, $10,000$ shots are taken for each iteration of the quantum and classical experiments. Each of the multi-shot classical MCMC experiments and MPS simulation of the quantum algorithm are repeated ten times. We also run at least three repeats of the converged quantum hardware experiments for three generations of the IBM quantum \textit{Heron} devices for each of the problem instances. Specifically, we use \textit{Heron r1} (\texttt{ibm\_torino}), \textit{Heron r2} (\texttt{ibm\_fez}) and \textit{Heron r3} (\texttt{ibm\_boston}), which has the lowest error rates of these devices. The relevant error data is available via the \href{https://quantum.cloud.ibm.com}{IBM Quantum platform}. Most slopes of the fitted lines in the right panel of Figure~\ref{fig:scaling_summary} are worse than those on the left panel. Yet, the gradient of the red line on the right panel of Figure~\ref{fig:scaling_summary}, representing the \textit{Heron r3} results from \texttt{ibm\_boston}, has a comparable gradient to the fitted lines visible on the left-hand side. These \textit{Heron r3} hardware results outperform both the equivalent classical MCMC test, and the simulated quantum results in this multi-shot setting. 
Interestingly, MPS simulations perform best for the $52$ node problem on the right-hand panel of Figure~\ref{fig:scaling_summary}. However, for the $80$ and $117$ node instances, the real \textit{Heron r3} hardware outperform MPS simulation performance. 
Together, these results suggest potential for a promising scaling property for the quantum algorithm with increasing problem size. Additional experiments will be needed to further support this conclusion, using additional problem instances.

\section{Conclusions} \label{sec:conclusion}


In this manuscript, we introduce a heuristic quantum optimization algorithm based on a QeMCMC sampling routine to solve non-trivial combinatorial optimization problem instances. 

We combine our QeMCMC implementation, incorporating a warm-started quantum optimization circuit that intentionally biases our proposal distribution towards high-quality solutions, with parallel tempering, commonly used to find ground states of many-body systems, to explore the available cost landscape more efficiently. Crucially, while our heuristic method does not offer provable convergence guarantees, we recover the global optimum for all considered problem instances with 17, 39, 52, 80 and 117 variables, using up to 117 qubits on real quantum hardware. The results suggest favorable scaling over the tested classical MCMC approaches. However, state-of-the-art classical approaches for these problems are not presented in baseline comparisons in Section~\ref{sec:classical_algo_results}, since the scale of the problems studied are easy to solve for the best classical solvers. This is evident from the CPLEX runtimes reported alongside the time-to-solution (TTS) analysis for the quantum experiments in Appendix~\ref{appendix:runtime_data}. As a next step, more experiments across larger sets of instances and additional problem classes are needed, including comparisons against dedicated MIS solvers at larger problem sizes, to strengthen our conclusions about the quantum algorithm. Alternative circuit constructions could further improve algorithm performance, and approaches for scaling to larger problems, such as problem decomposition or block sampling, should be investigated. Furthermore, more advanced classical MCMC approaches that exploit the relevant problem structure, including cluster and isoenergetic moves, should be tested. 

Nevertheless, heuristic quantum-enhanced sampling approaches such as the one explored here provide a direction for investigating the practical utility of quantum optimization before fault tolerance.

\subsection*{Note added regarding concurrent work}
During the final preparation of this manuscript, a closely related preprint appeared that proposes a similar overall idea including small-scale classical simulation \cite{Ferguson_QeSA_QePT_2026}. Our contribution differs in that we demonstrate the method experimentally on quantum hardware using over 100 qubits. We also introduce practical details and evaluate the concrete circuit designs and implementation-level improvements required to achieve scalable execution. 

\begin{acknowledgments}
    \emph{Acknowledgements} - The authors thank David Layden, Mario Motta, Lewis W. Anderson, Benjamin Jaderberg, Lev S. Bishop, Stephan Eidenbenz, Reuben Tate, Antonio Mezzacapo, Max Rossmannek, Stefano Mensa and Jason Crain for helpful discussions and insight. This work was supported by the Hartree National Centre for Digital Innovation, a collaboration between the Science and Technology Facilities Council and IBM. 
\end{acknowledgments}


\appendix

\begin{figure*}[t]
    \centering
    \includegraphics[width=\textwidth]{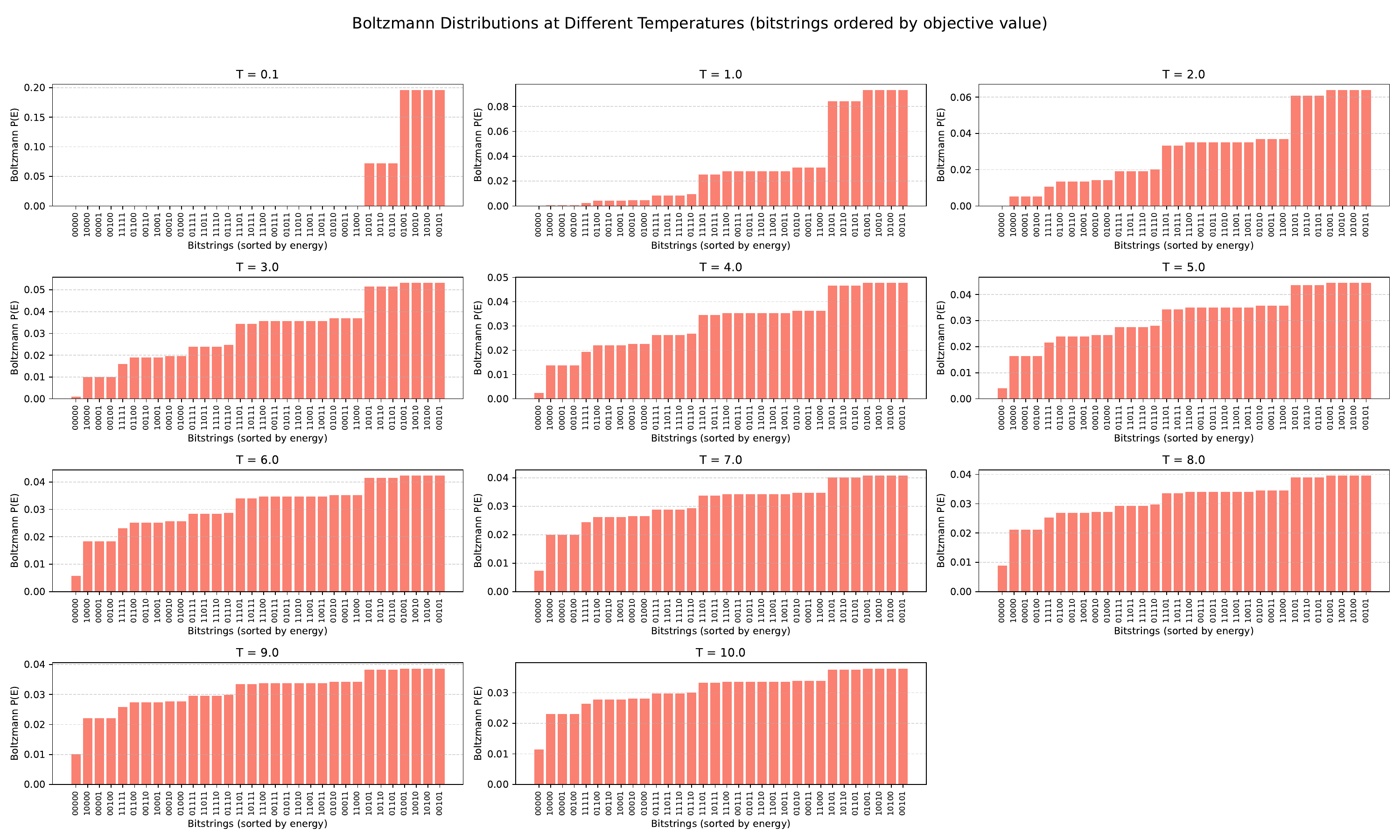}
    \caption{Theoretical distribution for a five node MIS problem instance, for a range of (dimensionless) temperatures $T$.}
    \label{fig:5_qubit_theoretical_distribution}
\end{figure*}

\begin{figure*}[t]
    \centering
    \begin{subfigure}[t]{\textwidth}
        \centering
        \includegraphics[width=\textwidth]{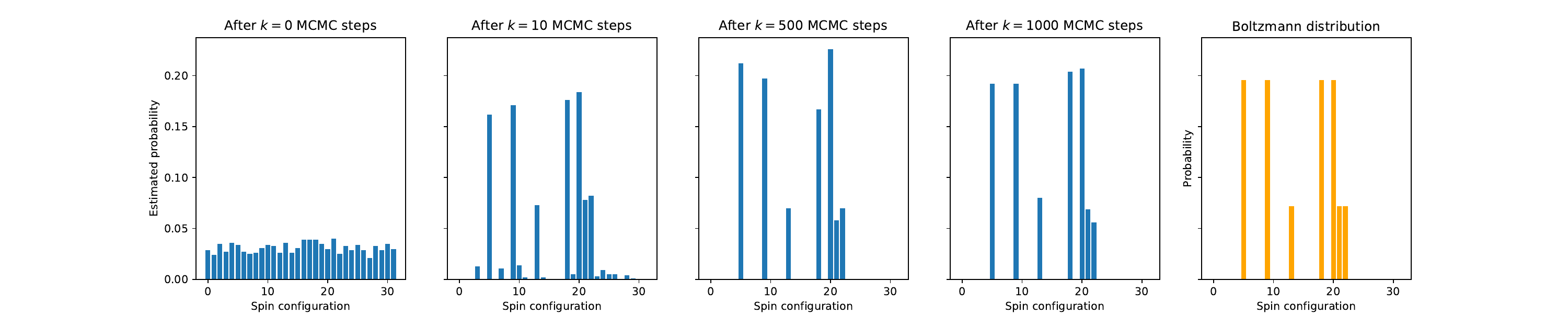}
        \caption{$T = 0.1$.}
        \label{fig:Boltzmann_low_T}
    \end{subfigure}
    \hfill
    \begin{subfigure}[t]{\textwidth}
        \centering
        \includegraphics[width=\textwidth]{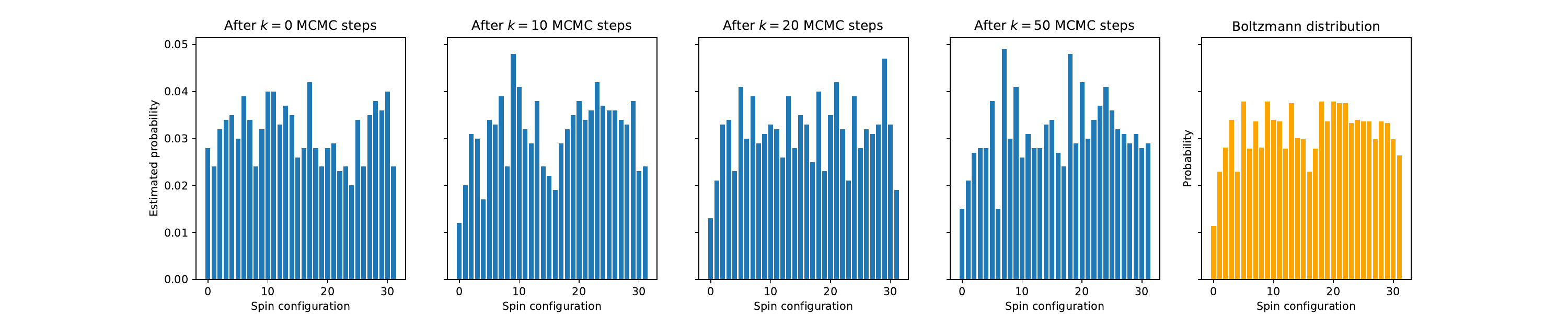}
        \caption{$T = 10$.}
        \label{fig:Boltzmann_high_T}
    \end{subfigure}
    \caption{Convergence to the stationary distribution for QeMCMC experiment at a high and low temperature.}
    \label{fig:boltzmann_sampling}
\end{figure*}

\section{Boltzmann sampling} \label{appendix:Boltzmann_sampling}

In the present manuscript, we implement circuits that resemble a first-order Trotter approximation of the discretized annealing Hamiltonian presented in Eq.~\eqref{eq:Hamiltonian}. This is as opposed to the second-order Trotter formulation used for circuits constructed in Ref.~\cite{QeMCMC_Layden_2023}. Using the first-order approximation is possible since we are not so concerned with Trotter error in the scope of this work. Additionally, the first-order Trotter approximation resembles QAOA, which is useful in our setup when scaling to $>100$ variable system sizes, as we can leverage advanced QAOA tools for optimizing parameters at these scales, such as MPS-based energy evaluators for classically tuning $\gamma$ and $\beta$ parameters. We demonstrate that with circuits resembling Eq.~\eqref{eq:qaoa}, our first-order Trotter decomposition of QeMCMC is still effective for sampling.

In preparation for using QeMCMC in the heuristic setting implemented in this paper, we conduct small-scale Boltzmann sampling experiments for various temperatures, as were performed in Ref.~\cite{QeMCMC_Layden_2023}. We do this using a small, five-node MIS problem instance. Owing to the modest dimensionality of this system, the complete distribution over computational basis states can be readily visualized. In Figure~\ref{fig:5_qubit_theoretical_distribution}, we see the theoretical Boltzmann distribution for this problem instance. Here, the Ising-like energies are computed by enumerating all $2^n$ bitstrings $\{s_i\}$ and evaluating their energies $E_i = E(s_i)$ under the MIS Hamiltonian. The corresponding probabilities are then obtained from the Boltzmann distribution defined in Eq.~(\ref{eq:boltzmann_distribution}), yielding the exact classical distribution at temperature $T$. 

In Figure~\ref{fig:Boltzmann_low_T}, we demonstrate the iterative process of converging to the desired Boltzmann distributions at $T = 0.1$, where $k_B = 1$ on IBM Quantum superconducting hardware. When Ref.~\cite{QeMCMC_Layden_2023} was originally published, the authors were limited from running these experiments end-to-end, within one execution window. This was due to limitations of running iterative experiments on the available quantum hardware, which is explained fully in the associated work \cite{QeMCMC_Layden_2023}. However, this execution style is now well-supported with \texttt{Qiskit Runtime} \href{https://quantum.cloud.ibm.com/docs/en/api/qiskit-ibm-runtime/session}{Session} mode. In the work presented here, we use Session mode alongside parameter streaming capabilities, which provide a tight classical feedback loop for determining circuit parameters based on quantum experimental results, within the Session window. The results shown in Figure~\ref{fig:Boltzmann_low_T} and Figure~\ref{fig:Boltzmann_high_T} were obtained in this way, which allows for efficient job batching and reduced queuing latency. These results are produced by first formulating the MIS problem for the five node problem using a Hamiltonian defined by the graph, to produce QAOA circuits of the form in Eq.~\eqref{eq:qaoa}, and then proceeding as per the logic described in Section~\ref{sec:qMCMC}. The QAOA circuits are used to construct a proposal, $Q$, explained in Section~\ref{sec:workflow}. We build $Q$ in the form of a dense matrix where each column is updated according to the starting state, $s$, and sampled solutions, $s'$, drawn from the quantum circuit. For a small problem like this, we can extract all components of this matrix exactly using brute force to build the full statevector, then enumerate until we have the full analytical matrix for $Q$. For larger problems, we would need to sample from quantum circuits to build up this matrix progressively, since the approach described here is resource intensive. We can also utilize sparse matrix approaches to approximate the matrix. Via both methods, we derive empirical probabilities of possible end states, $s'$, from the start state, $s$. 

Next, we apply the MH acceptance probability as defined in Eq.~\eqref{eq:accept_reject}, to construct a transition matrix, $P$, from the proposal matrix, $Q$, equivalent to $P$ in Eq.~\eqref{eq:transition_matrix}. The resulting transition matrix is stochastic, and fully describes our Markov chain, incorporating both the proposal distribution and the MH acceptance rule. Our transition matrix, $P$, is then used to generate moves in the Markov chain, and over many iterations, or steps, we can construct plots like that shown in Figure~\ref{fig:Boltzmann_low_T} and in Figure~\ref{fig:Boltzmann_high_T}. From this, we can check that our stationary distribution is an eigenvector of the transition matrix, $P$. 

We note that there is a general assumption within the community that HPC resources are required for implementing QeMCMC at non-trivial scales. In our work, HPC resources were used for a number of tasks both in experiment preparation and post-processing results, such as with parameter optimization and classical simulations of our largest implemented quantum circuits. However, generally, we avoided excessive use of HPC resources during the algorithm execution window by pre-emptively compiling circuits and using known QAOA best practices to scale our experiments. We also acknowledge that QeMCMC algorithms generally rely on a fast exchange of information between sampling from the quantum computer and a classical evaluation procedure to accept or reject a proposal. For the scales of problems investigated here, up to $117$ qubits, the use of remote APIs, such as the \texttt{Qiskit Runtime} \href{https://quantum.cloud.ibm.com/docs/en/api/qiskit-ibm-runtime/session}{Session} mode mentioned previously, were sufficient for our use case. However, we anticipate the classical computational demand of this algorithm to increase as we look to solve larger problem instances. 

\section{Maximum independent set mapping} \label{appendix:MIS_mapping}

We briefly explain the mapping from the QUBO defined in Eq.~\eqref{eq:lagrangian_form}, to a Hamiltonian of the form
\[
H_{\text{cost}} = H_{\text{objective}} + \lambda H_{\text{constraint}},
\]
up to a constant energy shift. We use a mapping with the relevant Paulis $x_i = (1 - Z_i)/2$, such that the MIS objective can be written as
\begin{equation}
    \sum_i x_i
    = \frac{|V|}{2} - \frac{1}{2}\sum_i Z_i
    = \frac{|V|}{2} + H_{\text{objective}},
    \label{eq:objective_hamiltonian}
\end{equation}
where $H_{\text{objective}} = -\tfrac{1}{2}\sum_i Z_i$, up to an additive constant.

Similarly, the constraint penalty becomes
\begin{equation}
\begin{aligned}
-\lambda \sum_{(v_i, v_j)\in E} x_i x_j
&= -\frac{\lambda}{4}\sum_{(v_i, v_j)\in E}
    \left(1 - Z_i - Z_j + Z_i Z_j\right) \\
&= -\frac{\lambda |E|}{4} + \lambda H_{\text{constraint}},
\end{aligned}
\label{eq:constraint_hamiltonian}
\end{equation}
where
\[
H_{\text{constraint}}
= \frac{1}{4}\sum_{(v_i,v_j)\in E}
\left(Z_i Z_j - Z_i - Z_j\right),
\]
again up to an additive constant. Here, the first term is a constant energy offset, while the remaining terms define the Hamiltonian encoding the constraints of the maximum independent set problem. These terms consist of single-qubit contributions proportional to $Z_i$ and pairwise interaction terms of the form $Z_i Z_j$. Upon promoting the classical spin variables to Pauli-$Z$ operators, the independent set objective and constraints take the form of a classical Ising Hamiltonian,
\begin{equation}
H_{\text{Ising}} = -\sum_i h_i Z_i \;-\; \sum_{(v_i,v_j)\in E} J_{ij} Z_i Z_j ,
\end{equation}
where the coefficients $h_i$ and $J_{ij}$ arise directly from the linear and quadratic operator terms in Eqs.~(\ref{eq:objective_hamiltonian}) and (\ref{eq:constraint_hamiltonian}), up to a similar additive constant. Finally, while the MIS problem is naturally posed as a maximization of the objective $\sum_i x_i$, we adopt the standard physics convention of defining an energy function as the negative of this objective (with additional penalty terms), thereby converting the problem into an equivalent energy minimization task whose ground states correspond to optimal solutions. 

\section{Circuit optimization} \label{appendix:circuit_opt}

In this Section, we discuss the tuning of parameters for individual QeMCMC replica circuits, which resemble $p=2$ QAOA. 
The MIS problem is mapped to a QUBO, described by an Ising Hamiltonian $H_{\text{cost}}$, by adding the constraints as penalty terms. The optimization landscape of our encoded problem changes with the value of this penalty term. As such, we tune this term in such a way that smoothens this landscape, generally with smaller penalty values.
The state $\psi({\beta}, {\gamma}; \lambda)$ constructed on the hardware thus depends on the QAOA angles $({\beta}, {\gamma})$ and the Lagrange multiplier $\lambda$. 
The latter is set by an optimization procedure that tries to equate the expectation value of the objective terms in $H_{\text{objective}}$ and the constraint terms of $H_{\text{constraint}}$ when sampling from $\ket{\psi}$, see details in Ref.~\cite{QOBLIB_Koch_2025}. 
This can result in good quality samples $s$ drawn from $\ket{\psi}$ when executed on quantum hardware, even though the ground state of $H_{\text{cost}}$ is not necessarily the same as the solution of the MIS problem when $\lambda<2$.

For each explored value of $\lambda$, the QAOA angles are optimized classically with the \textit{QAOA Training Pipeline}~\cite{qaoa_training_pipeline}.
This open-source package allows one to design a chain of parameter trainers to find ${\beta}$ and ${\gamma}$ in $\ket{\psi}$ along with the methods to evaluate the energy $\langle\psi|H_{\text{cost}}|\psi\rangle$.
To train parameters for our $p=2$ QAOA we first find optimal depth $p=1$ parameters with \texttt{SciPy} running COBYLA from the initial point $\beta_1=0$ and $\gamma_1=0$.
Next, the resulting optimal $\beta_1^\star$ and $\gamma_1^\star$ are used as the initial point for a transition state trainer to find the parameters at depth $p=2$. A constraint is applied such that $\gamma_1 = \gamma_2 \equiv \gamma$ and $\beta_1 = \beta_2 \equiv \beta$, as is discussed at the beginning of Section~\ref{sec:circuit_construction}. 

Here, we use $p=2$ QAOA, with three transition states. We optimize QAOA angles against the best energy, which is evaluated approximately with an MPS operating at various thresholds of $10^{-6}$ to $10^{-10}$, ensuring energies are converged with respect to truncation threshold~\cite{Sack2023}. 

\begin{figure}[h]
    \centering
    \includegraphics[width=\columnwidth]{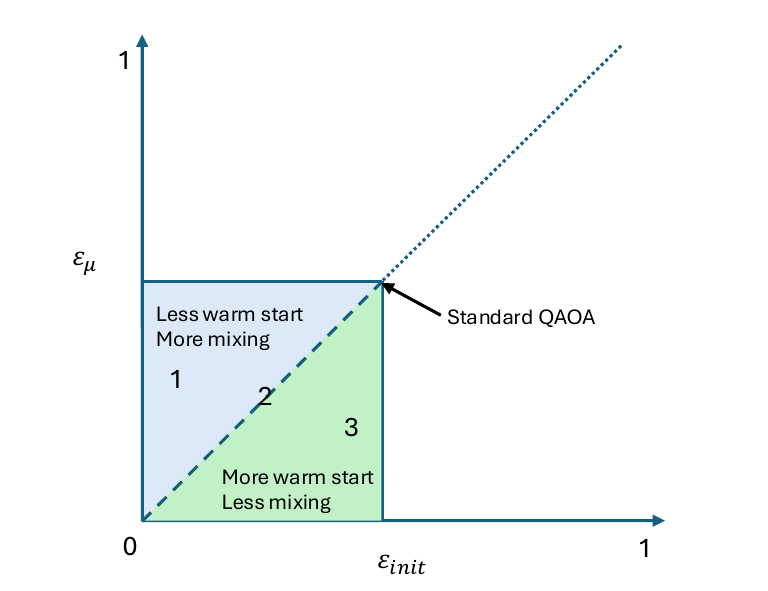}
    \caption{Visualization of different combinations of parameters $\varepsilon_{\text{init}}$ and $\varepsilon_{\mu}$. Regions labeled 1, 3 and 2, which indicate more or less warm starting and mixing respectively, were all tested in this work.}
    \label{fig:epsilon}
\end{figure}

Finally, to further enable our QAOA circuits which are used to generate the proposal distribution in our quantum-enhanced solver, we bias our circuits with a warm-starting implementation, as discussed in Section~\ref{sec:warm starting}. First, a relaxed version of the original problem is solved with a quantum circuit, and the resulting solution is used to warm-start the next iteration of our algorithm. This process is visualized in Figure~\ref{fig:workflow}. To do this, we extend the work documented in Ref.~\cite{WarmStarting_Egger_2021}. 

The process involves defining a regularization parameter, $\varepsilon$, which determines the rotation angles used in the initial state and mixer layer of subsequent QAOA circuits, as described in Section~\ref{sec:warm starting}. In this work, when tuning $\varepsilon$, we began testing with the same values originally used by authors in Ref.~\cite{WarmStarting_Egger_2021}. We also scanned nearby values to this starting point and tested for optimal performance. Finally, we also tried explicitly separating $\varepsilon$ for defining the $R_y(\theta)$ rotations applied to the initial state, $\varepsilon_{\text{init}}$, and for the mixer, $\varepsilon_{\mu}$, such that we could explore the continuum of options between partially and fully warm starting our circuits, or not at all. The different regimes we explored in the work documented in this paper are visualized in Figure~\ref{fig:epsilon}. However, we observed that the area of alignment between $\varepsilon_{\text{init}}$ and $\varepsilon_{\mu}$, i.e. on the diagonal of the plot, leads to the most performant results. This result agrees with conclusions made in Ref.~\cite{Aligment_He_2023}. With an aligned mixer constructed such that $\lvert s_\theta \rangle$ is its ground state, we can also expect to preserve the convergence guarantee in the limit $p \to \infty$ \cite{WarmStarting_Tate_2023}. 

\section{Heuristic implementation of QeMCMC} \label{appendix:heuristic_qemcmc}

\begin{figure*}[t]
    \centering
    \begin{subfigure}{0.7\textwidth}
        \centering
        \includegraphics[width=\textwidth]{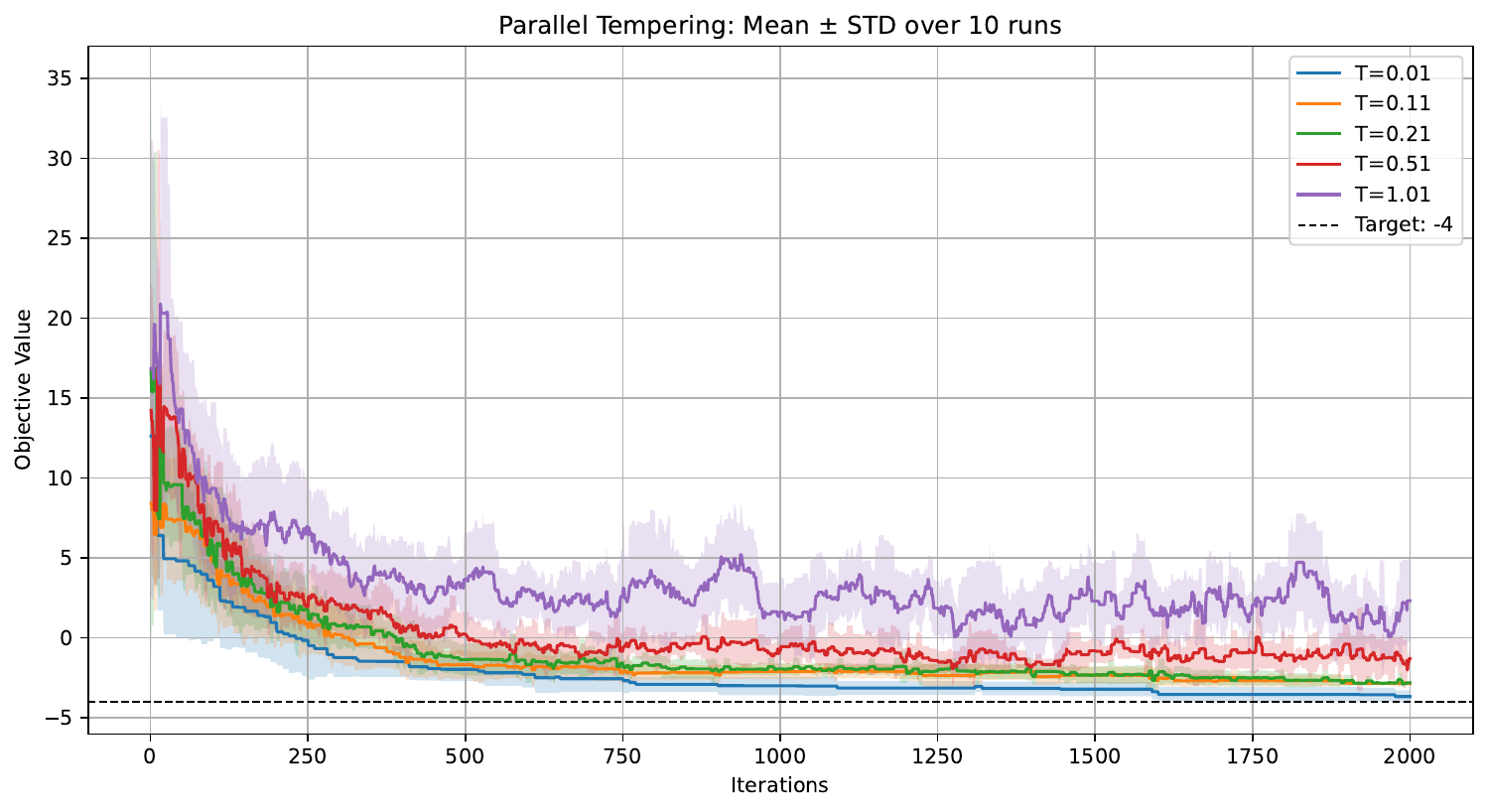}
        \label{fig:17_node_result_1_shot_QeMCMC_sim_warm_started}
    \end{subfigure}
    \hfill
    \begin{subfigure}{0.7\textwidth}
        \centering
        \includegraphics[width=\textwidth]{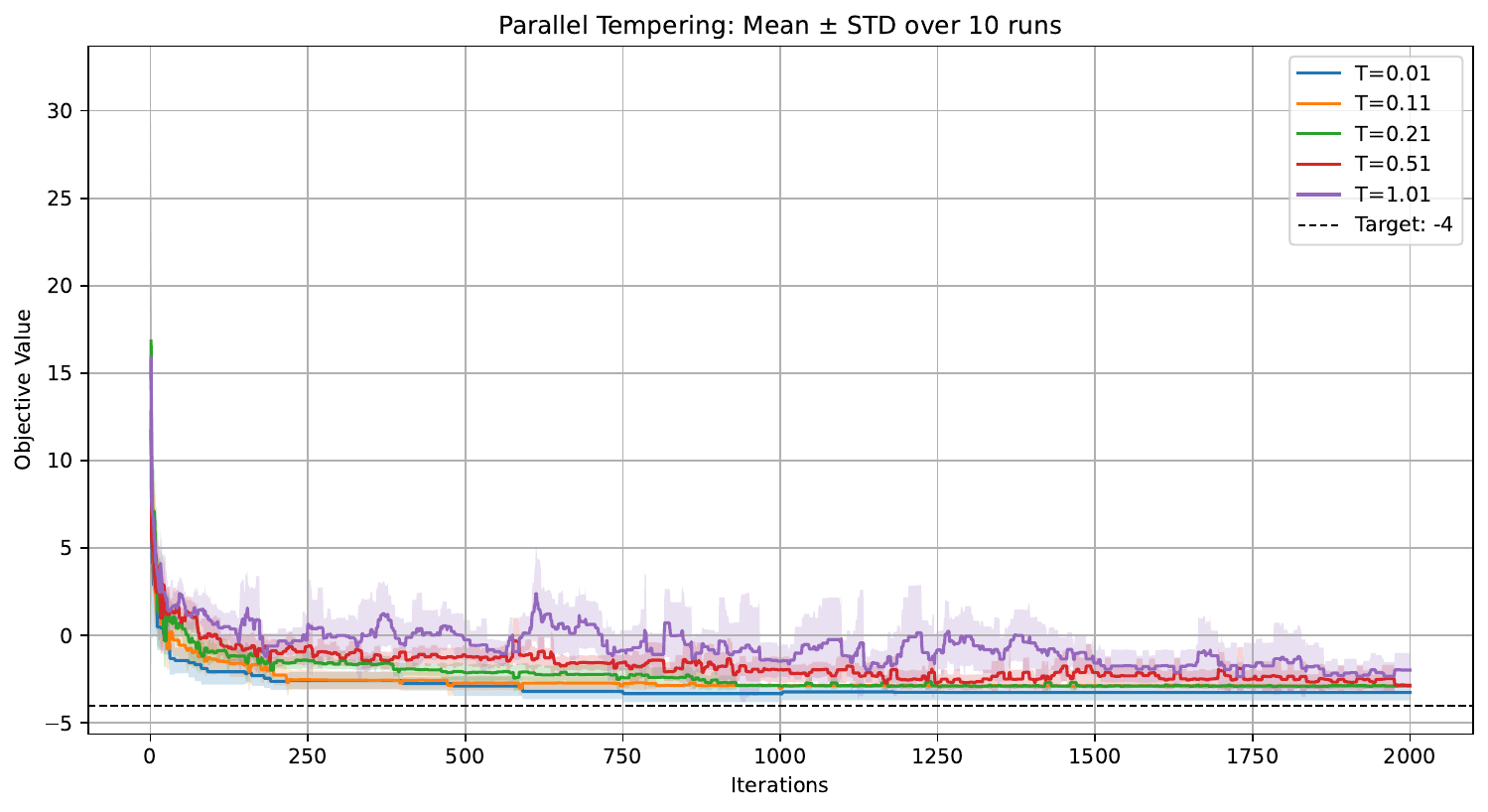}
        \label{fig:17_node_result_1_shot_QeMCMC_sim_not_warm_started}
    \end{subfigure}
    \caption{17 node graph MIS problem results, using an MPS-based sampling simulation of the quantum algorithm with (top panel) and without (bottom panel) detailed balance maintained. We see that we obtain similar convergence rates and behavior in both cases, when studying objective value against iterations. Each experiment is repeated 10 times, with 1 shot taken per iteration of the experiments, drawing a fair comparison between the two methods.}
    \label{fig:17_node_heuristic_comparison}
\end{figure*}

It is noted in the main text that due to our heuristic implementation of the QeMCMC method described in Section~\ref{sec:qMCMC}, we break the associated detailed balance condition and acceptance probability defined in Eq.~\eqref{eq:accept_reject}, and in turn, affect the overall search dynamics. Although we do not include a complete study of the implications of breaking detailed balance, we demonstrate in Figure~\ref{fig:17_node_heuristic_comparison} that for the smallest MIS problem instance of 17 nodes, the two variants exhibit comparable performance on this instance. The mean numbers of iterations to convergence are $1353.3$ when detailed balance is preserved and $1553.3$ when it is not. In this case, we simulate the experiments on simulated quantum hardware so as not to waste expensive quantum resources. We observe that, in practice, for the larger problem instances, breaking the detailed balance condition leads to more efficient algorithm performance. In fact, for the remaining four problem instances, we observed few and often no cases of successful convergence when detailed balance is preserved. This motivates our use of warm-starting methods and resampling for noise mitigation when testing the larger problem instances on quantum hardware. 

We note that our objective is to find the ground state of the cost Hamiltonian, rather than to obtain unbiased samples from a target Boltzmann distribution. We therefore use the Metropolis acceptance rule as a temperature-controlled heuristic for accepting or rejecting proposed solutions, analogous to its use in simulated annealing and parallel tempering for optimization. Since our proposal distribution is asymmetric, the resulting Markov chain is not guaranteed to sample from the target Boltzmann distribution. Nevertheless, this does not preclude its use as a heuristic optimization method.

\section{Tuning the parallel tempering experimental setup} \label{appendix:pt_setup}

\begin{figure*}[t]
    \centering
    \includegraphics[width=\textwidth]{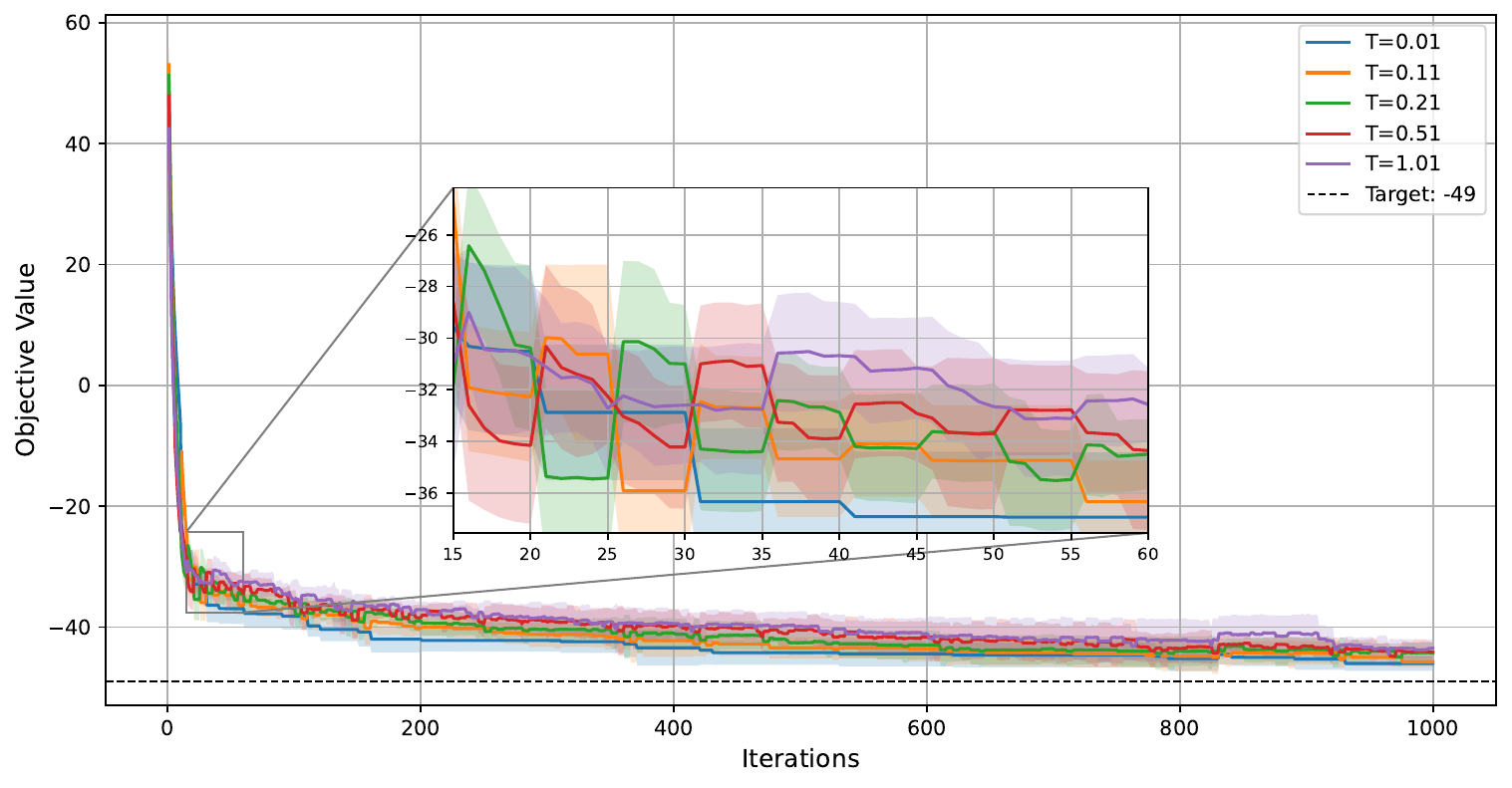}
    \caption{117 node graph MIS problem results, using an MPS-based sampling simulation of the quantum circuits. Iterations of the algorithm are plotted against the computed objective values, or energy, at each iteration. Each iteration involves taking $10,000$ samples, and randomly subsampling from a smaller subset of these samples. The results plotted are mean values with error bars computed from standard deviations, using ten repeats of the simulated tests. The inset shows faster mixing of individual replicas via replica exchange at regular intervals, towards the converged solution with an energy of $-49$.}
    \label{fig:tn_sim_117_nodes}
\end{figure*}

In this Appendix, we explain how the parameters used in the parallel tempering setup were selected. As a reminder, parallel tempering was initially developed to improve MCMC sampling of Gibbs distributions \cite{swendsen1986replica, geyer1991interface, hukushima1996exchange}, but it is also commonly adapted as an optimization tool to find low‑energy or ground‑state solutions, where we expect the target ground state energy to coincide with the expected value of the Gibbs free energy of a low temperature replica of the parallel tempering experiment \cite{earl2005parallel, wang2015comparing}. 

Accordingly, we use several standard design considerations from the parallel tempering literature \cite{earl2005parallel, swendsen1986replica, geyer1991interface, hukushima1996exchange, hansmann1997parallel, kone2005selection, lingenheil2009efficiency, nadler2008optimized, brooks2011handbook} to guide our experiments. Firstly, determining the number of replicas and their corresponding temperatures is key in ensuring algorithmic performance \cite{earl2005parallel}. In parallel tempering, the maximum temperature $T_{\text{high}}$ should be chosen so that replicas can escape local energy minima. The number of replicas must be sufficient to allow swaps to occur reliably between adjacent temperatures. In our setup, we determined the temperature range by first computing the maximum temperature using an inverted Boltzmann factor,
\begin{equation}
    T_{\text{high}} = \left| -\,\frac{x_{\text{max}}}{k_B \ln P_{\text{high}}} \right|, 
\end{equation}
where $x_{\text{max}}$ is the largest eigenvalue of $H_{\text{cost}}$, which is precomputed for our smaller problem instance of $17$ qubits using exact diagonalization. Since studying the spectrum of the Hamiltonian is only possible for our smallest problems, we rely on the fact that our problem instances are closely related in exhibiting heterogeneous degree distributions and nontrivial interaction structure. As such, we can appropriately extrapolate these values for the larger related problem instances, before tuning these values experimentally. $P_{\text{high}}$ is the target acceptance probability for high temperature moves, guided by the relevant literature which dictates $P_{\text{high}}\geq0.5$. The minimum temperature is a small, fixed value to allow adequate sampling of low-energy states. We further tune temperatures experimentally to  optimize performance on the hardware. 

Based on heuristic MCMC test runs, we clip $T_{\text{low}}  = 0.01$, to ensure that there is always a non-zero probability of a transition even in cases where proposed moves result in higher energy solutions. With this temperature range, number of replicas $N$ was chosen to achieve adequate swap acceptance between adjacent temperatures. For a fixed range $[T_{\text{low}}, T_{\text{high}}]$, we increased the number of replicas until adjacent temperatures yield satisfactory swap acceptance probabilities, typically in the range 20-40\% \cite{hansmann1997parallel, kone2005selection}. 

We then constructed a geometric temperature ladder 
\begin{equation}
    T_i \;=\; T_{\text{low}} \left( \frac{T_{\text{high}}}{T_{\text{low}}} \right)^{\frac{i-1}{N-1}},
\end{equation}
for $i=1,2,\dots,N$. This spacing concentrates replicas at lower temperatures (where energy barriers are more pronounced) while still permitting replicas with higher temperatures \cite{nadler2008optimized}.

Next, we consider the replica-exchange regularity, defined by the parameter $\Delta t_{\text{swap}} = 1/k$. Let $k$ denote the number of local MCMC updates performed between consecutive exchange attempts (so an exchange is attempted after every $k$ local updates). We selected $k$ experimentally to yield satisfactory swap acceptance probabilities. In our setup, $k$ was chosen in the range $1 \leq k \leq 5$. Furthermore, in the replica exchange stage of the algorithm, there exist several ways to implement the exchange step \cite{lingenheil2009efficiency}. In this work, we used a stochastic even-odd algorithm. Let
$
\mathcal{N}=\{(i, i+1) \mid i=1, \ldots, N-1\}
$
be the set of neighboring replica pairs in the temperature ladder. Furthermore, define the even and odd subsets:
$
\mathcal{E}=\{(2 j, 2 j+1)\}, \quad \mathcal{O}=\{(2 j-1,2 j)\}. 
$
At each iteration, we alternate between even and odd replica pairs, and swap attempts are performed for all pairs within the selected subset. This allows exchanges that propagate information down the temperature ladder toward the coldest chains. The algorithm then proceeds until termination, which in our setup occurs either when a target objective value is reached or when $200,000$ iterations are exhausted.

An example MPS simulation of the parallel tempering experiment run on hardware in Figure~\ref{fig:hardware_117_nodes} is presented in Figure~\ref{fig:tn_sim_117_nodes}, where the zoomed in panel allows us to visualize the replica exchange process at work. 

\section{QAOA ansatz construction} \label{appendix:ansaetze} 

QAOA composes an ansatz circuit following Eq.~\eqref{eq:qaoa}. Of course, the graph topologies of the problem instances used do not match the qubit connectivity graphs of the quantum hardware. Therefore, implementing the full cost operator requires deep networks of SWAP gates, even when advanced compilation and transpilation tools are employed, which themselves introduce various SAT mappings and SWAP strategies to try to reduce circuit depth \cite{qaoa_training_pipeline}. For example, the $117$ node problem would require a two-qubit gate depth of $696$ to fully implement $e^{-i\gamma H_{\text{cost}}}$. These deep circuits are problematic for noisy quantum hardware. Indeed, beyond a certain circuit depth, the gains of capturing more of the structure of the problem through additional SWAP gates is eroded by device noise. This tradeoff was explored in Ref.~\cite{Approx_Quadratization_Dragoi_2025} on random $3-$regular graphs, where best samples were obtained by a $p=2$ QAOA. This simplified ansatz $H_{\text{cost}}$ was chosen by allowing a limit to the number of allowed SWAP gates.

We construct simplifications of the problem that we can execute, by applying a SWAP network to the graph \cite{Weidenfeller_2022, Approx_Quadratization_Dragoi_2025}. The simplest instances are effectively hardware native graphs. We generate subgraphs of the problem that are overlapping, and become incrementally more complex (in terms of density of edges in the graph) with more SWAP layers allowed, resulting in a higher overlap with the full graph. From these subgraphs, we can select an ansatz circuit that sensibly balances quality of samples with feasibility to run on hardware. The QAOA ansatz that we implement is given by
\begin{align}
    U_{\text{QAOA}}=e^{-i\beta_2H_{\text{mix}}}e^{-i\gamma_2H_{\text{cost}}'}e^{-i\beta_1H_{\text{mix}}}e^{-i\gamma_1H_{\text{cost}}'}
\end{align}
where $H_{\text{cost}}'$ implements only the subset of the $Z_iZ_j$ terms that can be implemented with two layers of SWAP gates on a line of qubits. 

In choosing the ansatz for each of the problem instances, we study the behavior seen when increasing the number of SWAP layers to see whether this also increases the quality of the samples. We tune the quality of the QAOA circuit ansatz, to understand where sample quality no longer increase with the number of SWAP layers allowed, and after what point we see noise degrading the results. We observe how the quality of the ansatz performs for a $p = 2$ QAOA given a maximum number of SWAP layers,  $k$. We optimize angles for the QAOA for the chosen $(p, k)$ as per the process described in Appendix~\ref{appendix:circuit_opt}. Finally, we draw samples from the hardware and compute the average approximation ratio of the resulting objectives, for each of four possible ansaetz to be studied, in this example for the 52 node graph problem. Since we are working with constrained optimization, we filter for valid samples before computing their objective values and taking an average for the approximation ratio. We also test with and without a classical correction step for the proposed solutions to the MIS. This is a classical post-processing step that aggressively removes nodes across the proposals that violate the constraints for the MIS problem. The solution is then improved upon where possible. We apply this classical correction step in this analysis, since the full iterative algorithm proposed in this paper is not used and we are only testing the quality of samples from the ansaetz. It is noted that even classically corrected samples from the ansatz circuits tested without the iterative algorithm approach presented in this paper are not able to reach the true solution to this 52 node problem.

Using the results in Table~\ref{table:ansatz_analysis}, we see how number of SWAP layers, $k$, affects quality of the ansatz, informing our choice of ansatz for our algorithm. When we consider only feasible solutions to the raw circuit samples, we see quality of the samples (measured by the value of the approximation ratio between $[0,1]$) steadily increase up to $37$ SWAP layers, and then degrade. For those solutions that have been classically corrected, the approximation ratios are higher. The $37$ and $49$ SWAP layer ansaetz are most successful, since they have the highest overlap with the full graph. With this, we justify our decision to use the $37$ SWAP layer ansatz for this $52$ node problem.

\begin{table}[h]
\centering
\resizebox{\columnwidth}{!}{%
\begin{tabular}{|l|c|c|c|}
\hline
\textbf{Nodes/qubits} &
\textbf{No.\ SWAP layers} &
\begin{tabular}{c}
\textbf{Classical} \\
\textbf{post-processing}
\end{tabular} &
\textbf{Approx.\ ratio} \\
\hline
$52$ &
$13$ &
False &
$0.0172$\\
\hline
$52$ &
$25$ &
False &
$0.0241$\\
\hline
$52$ &
$37$ &
False &
$0.0403$\\
\hline
$52$ &
$49$ &
False &
$0.0394$\\
\hline
$52$ &
$13$ &
True &
$0.9074$\\
\hline
$52$ &
$25$ &
True &
$0.9002$\\
\hline
$52$ &
$37$ &
True &
$0.9109$\\
\hline
$52$ &
$49$ &
True &
$0.9111$\\
\hline
\end{tabular}%
}
\caption{Approximation ratios taken for various ansaetz to the full $52$ node graph, given a number of allowed SWAP layers.}
\label{table:ansatz_analysis}
\end{table}

\section{Quantum hardware runtime data} \label{appendix:runtime_data} 

\begin{table*}[!t]
\centering
\resizebox{0.75\textwidth}{!}{%
\begin{tabular}{|c|c|c|c|c|c|c|c|c|}
\hline

\begin{tabular}{c}
\textbf{Problem} \\
\textbf{instance} \\
\textbf{(nodes)}
\end{tabular}
&
\begin{tabular}{c}
\textbf{Num.} \\
\textbf{repeats}
\end{tabular}
&
\begin{tabular}{c}
$\mathbf{\hat{P}_{\mathrm{hit}}}$ \\
\textbf{(95\% CI)}
\end{tabular}
&
\begin{tabular}{c}
$\mathbf{\mathbb{E}[N]}$ \\
\textbf{trials}
\end{tabular}
&
\begin{tabular}{c}
$\mathbf{N_{95}}$ \\
\textbf{trials}
\end{tabular}
&
\begin{tabular}{c}
${t_{\mathrm{trial}}}$ \\
\textbf{(minutes)}
\end{tabular}
&
\begin{tabular}{c}
${\mathrm{TTS}_{95}}$ \\
\textbf{(minutes)}
\end{tabular}
&
\begin{tabular}{c}
${t_{\mathrm{solution,best}}}$ \\
\textbf{(minutes)}
\end{tabular}
&
\begin{tabular}{c}
${t_{\mathrm{CPLEX}}}$ \\
\textbf{(seconds)}
\end{tabular}
\\
\hline

17
& $10$
& $1.00\;[0.72,\,1.00]$
& $1.00$
& $1$
& $5.35$
& $5.35$
& $1.63$
& $0.0262$
\\

39
& $3$
& $1.00\;[0.44,\,1.00]$
& $1.00$
& $1$
& $21.63$
& $21.63$
& $11.86$
& $0.0343$
\\

52
& $3$
& $1.00\;[0.44,\,1.00]$
& $1.00$
& $1$
& $35.67$
& $35.67$
& $13.81$
& $0.0936$
\\

80
& $3$
& $0.67\;[0.21,\,0.94]$
& $1.50$
& $3$
& $121.81$
& $365.43$
& $85.85$
& $0.0497$
\\

117
& $10$
& $0.60\;[0.31,\,0.83]$
& $1.67$
& $4$
& $170.12$
& $680.48$
& $76.88$
& $1.0630$
\\

\hline
\end{tabular}
}

\caption{
Reliability and time-to-solution statistics for the heuristic quantum optimization algorithm executed on quantum hardware. We report the number of independent hardware experiments used to estimate the empirical single-trial success probability $\hat{P}_{\mathrm{hit}}$, for which a 95\% confidence interval (CI) is reported. $\mathbb{E}[N]$ is the expected number of independent trials required to obtain one successful solution, while $N_{95}$ is the minimum number of independent trials required to obtain at least one solution with 95\% probability. $t_{\mathrm{trial}}$ is the estimated runtime of a complete hardware trial without early termination, and $\mathrm{TTS}_{95}$ is the corresponding runtime for $N_{95}$ complete trials. $t_{\mathrm{solution,best}}$ is the shortest observed time at which an optimal solution was reached in a successful hardware experiment. $t_{\mathrm{CPLEX}}$ is the time required by CPLEX to solve the corresponding problem instance. Wall-clock times include quantum execution and classical processing time between iterations, but exclude time spent waiting in queues for hardware.}
\label{tab:runtime_data}
\end{table*}

In addition to the convergence scaling analysis presented in
Section~\ref{sec:comparison_scaling}, we analyze the reliability and wall-clock cost of the heuristic quantum optimization algorithm introduced in this manuscript, using the time-to-solution (TTS) statistics reported in Table~\ref{tab:runtime_data}. For each problem instance, the empirical single-trial probability of identifying an optimal solution is calculated as \[\hat{P}_{\mathrm{hit}}=\frac{N_{\mathrm{success}}}{N_{\mathrm{repeats}}},\] where $N_{\mathrm{success}}$ is the number of successful hardware experiments. Since only a small number of hardware experiments were feasible, a 95\% Wilson binomial confidence interval is reported for $\hat{P}_{\mathrm{hit}}$. This interval is preferred to the normal approximation for small sample sizes and for observed probabilities close to zero or one.

Assuming independent trials with constant success probability, the expected number of trials required to obtain the first successful solution is
\begin{equation}
    \mathbb{E}[N]
    =
    \frac{1}{\hat{P}_{\mathrm{hit}}}.
    \label{eq:expected_trials}
\end{equation}
The minimum number of independent trials required to obtain at least one successful solution with 95\% probability is
\begin{equation}
    N_{95}
    =
    \left\lceil
        \frac{\ln(0.05)}
             {\ln(1-\hat{P}_{\mathrm{hit}})}
    \right\rceil.
    \label{eq:n95}
\end{equation}
These values are calculated using the point estimates of the per-trial success probabilities. Accounting for uncertainty in these estimates can substantially increase the required number of trials; for example, using the lower 95\% Wilson confidence bounds gives approximately nine trials for the 117-node instance. For $\hat{P}_{\mathrm{hit}}=1$, $N_{95}=1$ directly. The estimated wall-clock duration of a complete trial run to the specified $200,000$ iteration cap, without early termination, is denoted $t_{\mathrm{trial}}$. The 95\% time-to-solution is then defined as
\begin{equation}
    \mathrm{TTS}_{95}
    =
    N_{95}\,t_{\mathrm{trial}}.
    \label{eq:tts95}
\end{equation}
We also report $t_{\mathrm{solution,best}}$, which is the shortest observed time at which an optimal solution was reached and therefore represents the best observed successful runtime. 

For reference, the CPLEX solution time is also reported, which makes evident that the studied problem instances here are easy to solve for advanced classical solvers.

\bibliography{refs}

\end{document}